\documentclass[a4paper,11pt]{article}

\pdfoutput=1

\usepackage{jcappub}

\usepackage{color} 
\definecolor{Blue}{rgb}{0,0,0.9}
\definecolor{Red}{rgb}{0,0,0.9}
\definecolor{Green}{rgb}{0,0,0.9}
\definecolor{orange}{rgb}{0.8, 0.33, 0.0}

\usepackage[T1]{fontenc}
\usepackage{widetext}
\usepackage{subfig}
\usepackage{booktabs}

\title{\boldmath Relation between the Turnaround radius and virial mass in $f(R)$ model}

\author[a,b]{Rafael C. C. Lopes,}
\author[a]{Rodrigo Voivodic,}
\author[a]{L. Raul Abramo,}
\author[c]{Laerte Sodr\'{e} Jr}

\affiliation[a]{Instituto de F\'isica, Universidade de S\~ao Paulo, Rua do Mat\~ao, 1371, S\~ao Paulo, SP 05508-090, Brazil}
\affiliation[b]{Instituto Federal de Educa\c{c}\~{a}o, Ci\^{e}ncia 
e Tecnologia do Maranh\~{a}o, Campus Santa In\^es, BR-316, s/n, Santa In\^{e}s, MA 65304-770, Brazil }
\affiliation[c]{Instituto de Astronomia, Geof\'{i}sica e Ci\^{e}ncias Atmosf\'{e}ricas, Universidade de S\~{a}o Paulo, Rua do Mat\~{a}o, 1226, S\~{a}o Paulo, SP 05508-090, Brazil}

\emailAdd{rafaelcastro@ifma.edu.br}
\emailAdd{rodrigo.voivodic@usp.br}
\emailAdd{abramo@fma.if.usp.br}
\emailAdd{laerte.sodre@iag.usp.br}

\abstract{We investigate the relationship between the turnaround radius 
$R_t$ and the virial mass $M_v$ of cosmic structures in the context of $\Lambda$CDM model and in an $f(R)$ model of modified gravity -- namely, the Hu-Sawicki model.
The turnaround radius is the distance from the center of the cosmic structure to the shell that is detaching from the Hubble flow at a given time, while the virial mass is defined, for this work, as the mass enclosed within the volume where the density is $200$ times the background density. 
We employ a new approach by considering that, on average, gravitationally bound astrophysical systems (e.g., galaxies, groups and clusters of galaxies) follow, in their innermost region, a Navarro-Frenk-White density profile, while beyond the virial radius ($R_v$) the profile is well approximated by the 2-halo term of the matter correlation function. By combining these two properties together with the information drawn from solving the spherical collapse for the structures, we are able to connect two observables that can be readily measured in cosmic structures: the turnaround radius and the virial mass. 
In particular, we show that, in $\Lambda$CDM, the turnaround mass at $z=0$ is related to the virial mass of that same structure by $M_t \simeq 3.07 \, M_v$, while in terms of the radii we have that $R_t \simeq 3.7 \, R_v$ (for virial masses of $10^{13} \, h^{-1} \, M_\odot$). 
In the $f(R)$ model, on the other hand, we have $M_t \simeq 3.43 \, M_v$ and $R_t \simeq 4.1 \, R_v$, for $|f_{R0}|=10^{-6}$ and the same mass scale. 
Therefore, the difference between $\Lambda$CDM and $f(R)$ in terms of these observable relations is of order $\sim 10-20\%$ even for a relatively mild strength of the modification of gravity ($|f_{R0}|=10^{-6}$).
For the turnaround radius itself we find a difference of $\sim 9\%$ between the weakly modification in gravity considered in this work ($|f_{R0}|=10^{-6}$) and $\Lambda$CDM for a mass of $10^{13} \, h^{-1} \, M_\odot$.  
Once observations allow precisions of this order or better in measurements of the turnaround $R_t$, as well as the virial mass $M_v$ (and/or the virial radius $R_v$), these quantities will become powerful tests of modified gravity.}


\begin{document}
\maketitle
\flushbottom

\section{Introduction}
\label{sec:intro}
The unknown cause of the accelerated expansion of the universe, observed now for more than 20 years \cite{riess1998, perlmutter1999}, is one of the most pressing open issues in cosmology.
The standard explanation introduces a new energy density component called Dark Energy (DE) \cite{amendola2010dark}, and out of the possible types of dark energy, the cosmological constant ($\Lambda$) is the simplest choice.
However, even if the $\Lambda$CDM model seems consistent with all observations, 
the small value of the cosmological constant does not sit well with the rest of physics \cite{weinberg1989cosmological}, and this is one of the motivations to
investigate other possible causes for the accelerated expansion. 
One of the alternatives is to tweak the standard theory of gravity, General Relativity (GR), and replace it with some theory of modified gravity (MG) \cite{carroll2004cosmic}. The question, then, becomes how to distinguish these theories from $\Lambda$CDM.

Here we consider the $f(R)$ class of MG models, specifically the Hu-Sawicki (HS) model \cite{sawicki2007stability,hu2007models}. In this class of models the Ricci scalar 
$R$ is replaced, in the Einstein-Hilbert action, by $R+f(R)$. 
It has been studied in many contexts, e.g., voids \citep{voivodic2016modelling}, spherical collapse model \cite{PhysRevD.85.063518, PhysRevD.88.084015, Chakrabarti2014, cembranos2012gravitational, herrera2017calculation}, structure formation \cite{brax2012structure}, the matter power spectrum \citep{oyaizu2008nonlinear}, in the formation and evolution of stars \cite{Capozziello:2011gm}, in clusters of galaxies \cite{PhysRevD.92.044009} and in type Ia supernovae \cite{PhysRevD.93.084016}.
However, our interest here is in the event of the turnaround, when an overdensity, after expanding together with the Hubble flow, reaches a maximum size, and subsequently collapses.

The turnaround radius, $R_t$, defined here as the distance from the center of one such structure up to the surface of null radial velocity, has been studied as a viable cosmological test in, e.g., \citep{Pavlidou:2014aia, Pavlidou:2013zha, Tanoglidis:2014lea,Tanoglidis:2016lrj}.
Recently, \citep{nojiri2018effects} also explored the effects of including 
$R^2$ corrections to gravity on $R_t$, claiming deviations of the order of $10\%$ from GR.

Measurements of $R_t$ can be used, in particular, as a test of the $\Lambda$CDM model, because in that scenario the turnaround radius has an upper limit $R_{t,max}=\left(\frac{3GM}{\Lambda c^2}\right)^{1/3}$ \citep{Tanoglidis:2014lea}. 
This new observable is useful for testing the laws gravity on cosmological scales, as data on the velocity profiles of groups and clusters can be used to compare the turnaround radius with theoretical predictions \citep{Pavlidou:2014aia}. 
Moreover, the turnaround moment is of particular interest for tests of MG because it relates to mechanisms taking place in low density regions, where screening mechanisms are expected to be sub-dominant, or at least to act only in a weak sense \cite{khoury2004chameleon}.
This perspective has attracted even greater attention after recent claims about apparent violations of the $R_{t, max}$ limit \citep{lee2015bound, 2017arXiv170906903L}, and after it was shown that modifications in the laws of gravity can affect this maximum turnaround radius \citep{Bhattacharya2017cosmic, lee2017effect, bhattacharya2017maximum}.

We should emphasize that these upper bounds for $R_t$ were derived using a definition of turnaround which employs the null {\it acceleration} surface ($\left. \ddot{r} \right|_{R_t}=0$), whereas the actual observable is the turnaround radius $R_t$ defined in terms of velocity, $\left. \dot{r} \right|_{R_t}=0$.
Focusing on $R_t$ defined in this way, we have studied the properties of turnaround in the context of $\Lambda$CDM and in the HS model \citep{lopes2018turnaround}.
Our results show that the density contrast at the turnaround moment, $\delta_t$, can decrease by $\simeq 18\%$, and the turnaround radius can increase by $\simeq 6.4\%$, at redshift $z=0$ for a mass of $10^{13} \, h^{-1} \, M_{\odot}$, with the MG parameter $f_{R0}=10^{-6}$ (these are, respectively, the optimal scale of mass for comparing 
$R_t$ measurements \citep{tanoglidis2015testing}, and the weakest value of the modified gravity parameter considered in Ref. \citep{lopes2018turnaround}).
We also computed the dependence of $R_t$ with redshift and with the turnaround mass $M_t$ (the mass within $R_t$), and showed that measurements of this observable over a range of mass scales and redshifts could distinguish the HS model from the $\Lambda$CDM model. 

However, the relation between  $R_t$ and $M_t$ is very difficult to measure -- especially since a direct measurement of $M_t$ seems unattainable.
The process of collapse happens shell by shell: by the time the innermost shells are close to virialization, an outer shell will be experiencing turnaround. 
In fact, for any given structure where we detect the place and time of turnaround, the corresponding mass which can be most easily observed is not $M_t$, but rather the mass of the region which is closest to virialization. 
Here we characterize this virialized mass $M_v$ in terms of a certain density of $\Delta$ times the background density -- and in that respect we follow the literature, taking $\Delta = 200$. 
Hence, the goal of this paper is to bridge the gap between calculations in $\Lambda$CDM and MG and actual observations of cosmic structures, by relating the turnaround radius $R_t$ with the virial mass $M_v$.

This work is organized as follows: 
In Section \ref{sec:1} we review the main equations describing gravitational collapse in the context of theories of modified gravity. 
In Section \ref{sec:2} we show how to solve the spherical collapse model, and how to detect the time, radius and density contrast at the turnaround moment.
Then, in Section \ref{sec:3} we turn to a characterization of the density profile in terms of the halo model with the sum of two contributions: we take the density inside the central halo to follow the Navarro-Frenk-White profile, while the outer regions (up to turnaround) are assumed to follow the linear halo-matter correlation function.
In Section \ref{sec:4} we build a relation between the two shells, one which corresponds to the virialized region, and the other which lies at the surface of turnaround. 
Finally, with this results in hand, we are able to express, in Section \ref{sec:5}, the precise relation between the turnaround radius and the virial mass, $R_t(M_v)$, including the effects of MG. We also show, in this section, a comparison between our results and some current observations with some comments about the future improvements in theory and data. We summarize our results and main conclusions in Section \ref{sec:conc} and, in appendix \ref{append}, we describe the effects of modified gravity in the density profile.

\section{Modified gravity equations and $f(R)$ models}
\label{sec:1}

The class of models of modified gravity $f(R)$ proposes a modification in the Einstein-Hilbert action as 
\begin{equation}
S=\frac{1}{2 \kappa}\int d^{4}x  \sqrt{-g} \, [R+f(R)] 
\, ,
\label{eq:acaoEH}
\end{equation} 
where $\kappa=8 \pi G$ 
(for a review see, e.g., \cite{sotiriou2010f}). 
In particular, by replacing $f(R) \to -2\Lambda$ one recovers the $\Lambda$CDM model.

The field equations which follow from Eq. \eqref{eq:acaoEH} are
\begin{equation}
G_{\mu \nu} + f_R R_{\mu \nu} - \left(\frac{f}{2} - \Box f_R \right)g_{\mu \nu}- \nabla_{\mu} \nabla_{\nu} f_R = 
8\pi G T_{\mu \nu} \, ,
\label{eq:EqEinstModi}
\end{equation}
where $f_{R} \equiv df/dR$, and the Einstein and energy-momentum tensors are $G_{\mu \nu}$ and $T_{\mu \nu}$, respectively.
From the trace of Eq. \eqref{eq:EqEinstModi} we obtain a 
Klein-Gordon equation for the effective scalar field $f_R$, 
which can be regarded as an extra degree of freedom in $f(R)$ models,

\begin{equation}
3 \Box f_{R} - R + f_{R}R - 2f = -8\pi G T \,, \label{eqfieldfR}
\end{equation}
where $T$ is the trace of the energy-momentum tensor. 
From this equation it can be seen that the scalar field has an mass and effective potential given by
\begin{equation}
m_{f_R}^{2} = \frac{\partial^2 \, V_{\rm eff}}{\partial f_R^2} = 
\frac{1}{3} \left( \frac{1 + f_R}{f_{RR}} - R \right)
\quad \quad  \; \mathrm{and} \quad \quad
\frac{\partial V_{\rm eff}}{\partial f_R} 
\equiv \frac{1}{3} \left[ R- f_R R + 2 f - 8\pi G \rho \right] \,,   \label{potn}
\end{equation}
where we have assumed a matter-dominated Universe.
A simple expression for $m_{f_R}$ follows by considering the approximation $|Rf_{RR}|\cong f_R <<1$, and hence $m^2_{f_R}\approx 1/(3 f_{RR})$. 

This effective mass defines a range for the scalar field interactions through the associated Compton wavelenght, $\lambda_c \equiv \frac{1}{m_{f_R}}$, and as such this interaction can be regarded as a fifth force. 
The mechanism responsible for enhancing the mass of the scalar in dense regions of the Universe, effectively eliminating this fifth force on small scales, is known as "the chameleon mechanism" \cite{khoury2004chameleon}. 

In the "quasistatic" regime, 
the equations that describe the evolution of the curvature potential $\Psi$ (the spatial component of the scalar metric perturbations) are
\begin{equation}
\nabla^2 \delta f_{R}=\frac{a^2}{3}\left[\delta R  -8\pi G \delta \rho_m  \right] \; , \quad \quad  \mathrm{and} \quad \quad
\nabla^2 \Psi = \frac{16 \pi G}{3}a^{2} \delta \rho_{\rm m} - \frac{a^{2}}{6} \delta R(f_R) \, , \label{eq:eqpoissontofrandpotorig}
\end{equation}
where $\delta f_{R} = f_{R}(R)-f_{R}(\bar{R})$, $\delta R = R-\bar{R}$, and $\delta \rho_m = \rho_m-\bar{\rho}_m$ (bars represent spatial averages).
From these equations, combined with the spatial components of the Einstein equations, we can express $\Psi$ in terms of the Newtonian potential $\Phi$, leading to a modified linearized Poisson equation in Fourier space:
\begin{equation}
k^2 \Phi({\bf k})  = - 4\pi G \left[ 1+ \epsilon(k,a) \right]
a^2 \delta \rho_{\rm m}({\bf k}) \; , 
\quad \quad \mathrm{with} \quad \quad 
\epsilon(k,a)\equiv \frac{1}{3} 
\left(1+ \frac{a^{2} \, m_{f_R}^{2}(a)}{k^2} \right)^{-1} \; ,
\label{eq:poissoneqfrlinear}
\end{equation}   
where $\epsilon$ parametrizes the modifications of GR at linear order. 
There are two limiting cases which describe the maximum and the absence of modifications of GR: the first case occurs when the scales of interest $\lambda$ are much smaller than the Compton wavelength $\lambda_c$, i.e., $\lambda \ll \lambda_c$, or $a^2 m_{f_R}^2 \ll k^2$, resulting in $\epsilon \to 1/3$. This is known as large-field limit. The other case happens when $\lambda \gg \lambda_c$ or, equivalently, $a^2 m_{f_R}^2 \gg k^2$, and this is called the small-field limit, in which $\epsilon \to 0$ and we recover GR.

In this paper we have chosen to work with the MG model proposed by Hu \& Sawicki \citep{hu2007models}, where
\begin{equation}
f(R) = -m^2 \frac{c_{1}(R/m^2)^n}{c_{2}(R/m^2)^n + 1} \, ,
\label{frhusawick0}
\end{equation}
with $m^2 \equiv 8 \pi G \bar{\rho}_{\rm M}/{3}$ being a characteristic  mass scale of the model, and $c_1$, $c_2$ and $n$ dimensionless free parameters.
In the high curvature regime $R/m^2 \gg 1$, and assuming $n=1$, this model reduces to a cosmological constant plus an $1/R$ correction,
\begin{equation}
f(R) = -16 \pi G \rho_\Lambda - \bar{f}_{R0} \frac{\bar R_0^2}{ R} \, ,
\label{fRapprox}
\end{equation}
where $f_{R0}$ corresponds to the background scalar field today, $\bar{R}_0$ is the present value of the Ricci curvature in the background, and $\bar{f}_{R0}=f_R(\bar{R}_0)$.
The mass of the effective scalar field can be written, following the prescription of \cite{brax2012structure}, as
\begin{equation}
m_{f_R}(a)= m_{0} \left(
\frac{\Omega_{m0} \, a^{-3} + 4 \, \Omega_{\Lambda 0}}{\Omega_{m0} + 4 \,  \Omega_{\Lambda 0}} \right)^{3/2} \, ,
\label{eqma}
\end{equation}
with $m_{0}= {H_0}
\sqrt{(\Omega_{m0} +4 \, \Omega_{\Lambda 0}) / 2 |f_{R0}|}$.
In our calculations we have used the cosmological parameters
$\Omega_{m0}=0.31$, $\Omega_{\Lambda 0}=0.69$, $\sigma_8=0.86$ and $n_s=0.96$. 

\section{Spherical collapse in \texorpdfstring{$f(R)$}{f(R)}}
\label{sec:2}
Combining the Euler equation and the continuity equation for an irrotational pressureless perfect fluid of nonrelativistic matter with spherical symmetry, we obtain the nonlinear equation for the density contrast
\begin{equation}
\delta''+\left({3\over a}+{H'(a)\over H(a)}\right)\delta'-{4{\delta'^{2}}\over {3(1+\delta)}}=\frac{1+\delta}{H(a)^2 a^4} \nabla^2 \Phi
\, ,
\label{deltanonlinear}
\end{equation}
where $'=d/da$ and $E = H/H_{0}$.
Following \citep{brax2012structure}, the term $\nabla ^{2}\Phi$  is obtained from the linearized Poisson equation 
\begin{equation}
\nabla ^2 \Phi(\vec{x}, a)= \frac{3 H_{0}^{2}\Omega _{m}}{4 \pi ^{2}} \frac{1}{a} \int _{0}^{\infty}dk \left[1+\epsilon (k, a)\right]\delta (\vec{k}, a) \frac{k}{x}\sin (kx) \,.
\label{poissonequation}
\end{equation}

Given an initial density profile we can solve Eqs. \ref{deltanonlinear}-\ref{poissonequation}, tracing the spherical collapse from beginning to end. In particular, we are able to determine the scale factor at the moment of turnaround for each shell, $a_t$, the density contrast corresponding to the matter inside that shell, $\delta_t=\delta(a_t)$, as well as the turnaround radius $R_t$ itself, of course. 
In order to assess a wider variety of real density profiles we consider two idealized scenarios: a hyperbolic tangent profile (Tanh), and a ``physical'' density profile (Phy) -- for details see \cite{lopes2018turnaround}. 
In this approach $R_t$ can be written as  
\begin{equation}
R_t(a, M_{v})= a\left[ \frac{3 M_{t}(R_t, M_{v})}{4 \pi \Omega_{m0} \rho_{c}(1 + \delta _{t}[a, M_{t}(R_t, M_{v})])} \right]^{1/3} \, ,
\label{eq:turnradius}
\end{equation}
where $M_{t}$ is the mass enclosed by $R_{t}$ and  $M_v$ represents the virial mass. We also used $\rho_m(a)=\bar{\rho}_m(a)[1+\delta(a, M_v)]$, with $\bar{\rho}_m (a)=\rho_{m0} \, a^{-3}$ and $\rho_c=2.77 \, h^{-1} \, M_{\odot}/(h^{-1} {\rm Mpc})^3$.
It should be noted that the density contrast $\delta _{t}$ is a function of $M_{t}$ as well.

The main results of this approach were presented in \cite{lopes2018turnaround}, but in that paper we expressed the turnaround radius as a function of turnaround mass, not in terms of the virial mass. While the two should be related, the virial mass is much more accessible to observations.
However, in order to construct a relation between $M_v$ and $M_t$, we need to establish a density profile which can describe the distribution of mass from the central regions of the structure all the way to the turnaround radius.

\section{Density profile}
\label{sec:3}

The turnaround radius can be measured, in principle, through observations of the infall pattern of galaxies around groups and clusters of galaxies.
The virial mass, on the other hand, can be measured in a variety of ways, from richness to X-rays to the thermal Sunyaev-Zel'dovich effect, as well as gravitational lensing. 
Our aim in this Section is to specify the density profile of the inner and outer regions, in a framework that allows us to consider not only $\Lambda$CDM but also MG models \citep{beraldo2013testing}.

\subsection{NFW profile}
In order to relate the turnaround mass with the virial mass of self-gravitating systems, we start from the central regions of cosmic structures, which in a variety of scenarios seems to be well approximated by the density profile proposed by Navarro et al. \citep{Navarro:1995iw, navarro1997universal}. This profile, obtained by fitting stacked matter distributions of halos in cold dark matter (CDM) N-body simulations, is given by
\begin{equation}
\rho(r)=\frac{\rho_s}{(r/r_s)(1+r/r_s)^2},
\label{eqNFW}
\end{equation}
where $\rho_s$ and $r_s$ are scale parameters. The total mass inside a radius $R_{max}$ can be written as
\begin{equation}
M(<R_{max}) =\int_0^{R_{max}} 4\pi r^2 \rho_{NFW}dr=4 \pi \rho_s r_s^3\left[\ln\left(\frac{r_s+R_{max}}{r_s}\right)-\frac{R_{max}}{r_s+R_{max}}\right]\,.
\label{eqmassmax1}
\end{equation}
The scale radius $r_s$, in turn, can be expressed in terms of the virial radius $R_v$, which is related to the concentration parameter $c$ as $r_s=R_v/c(z, M_v)$. 
In this work we use the concentration fitted by \cite{bullock2001profiles}, $c(z, M_v)= 9 \, a \, ({M_v}/{M^*})^{-0.13}$, where $M^*$ is the mass for which $\sigma(M^*)=\delta_c$ (the critical linear density contrast at collapse, which in the Einstein-de Sitter model is $\delta_c = 1.686$). Therefore, in terms of $M_v$, we can rewrite the total mass inside a radius $R_{max}$ as
\begin{equation}
M(R_{max}, M_v, z)=4 \pi \rho_s \left[ \frac{R_{v}}{c(z, M_{v})}\right]^3
\left[\ln\left(1+c(z, M_{v}) \frac{R_{max}}{R_v} \right)-\frac{c(z, M_{v})}{c(z, M_{v})+\frac{R_v}{R_{max}}}\right].
\label{eq:massmax2}
\end{equation}

\subsection{Halo model}
Beyond the virial radius $R_v$, the matter density should asymptote to the mean density of the Universe at the given redshift -- in fact it is well known that
the NFW profile is not a good fit for radii larger than  $R_v$ \citep{beraldo2013testing}. However, since by assumption there is a virialized halo at the center (with mass $M_v$), the matter distribution around such halos should be given in terms of the matter correlation function -- more precisely, by the halo-matter cross-correlation, $\xi_{hm}(r)$. 

Hence, in this work we employ the halo model description, which treats the dark matter halos as the fundamental building blocks of the gravitational structures in the universe \cite{cooray2002halo}.
The halo-matter correlation function, which describes the excess of mass at a distance $r$ from the center of a halo, is $\xi_{hm}(r)=\langle\delta_h(\mathbf{x})\delta_m(\mathbf{x} + \mathbf{r})\rangle$, where $r=|\mathbf{r}|$ is the distance from the center. 
The correlation function $\xi_{hm}(r)$ is, in fact, an average of the observed overdensity around halos of similar masses, $\langle \rho_{h, obs}(r)\rangle$, since $\xi_{hm}(r)=\langle\rho_{h, obs}(r)\rangle/\bar{\rho}_m-1$.
In the context of the halo model, the average observed overdensity is given by the sum of two contributions \cite{hayashi2008understanding,schmidt2009nonlinear,beraldo2013testing, cooray2002halo},
\begin{equation}
\rho_{h,obs}(r)=\rho_{1h}(r)+\rho_{2h}(r),
\label{eq:density}
\end{equation}
where we have employed $\rho_{1h}(r)=\rho_{NFW}$ for the one-halo term that describes the contribution of the halo itself for the density profile, while the second term, called of two-halo term, represents the contribution from the large-scale structures exterior to the halo. The two-halo term can be written as 
\begin{equation}
\rho_{2h}(r)=\bar{\rho}_{m} \, b^L(M_h) \, \xi^L_m(r),
\label{eq:2haloterm}
\end{equation}
where $b^L(M_h)$ is the linear halo bias for halos of mass $M_h$, which we have modeled using the fit of Tinker \textit{et al.} \footnote{In the modified gravity cases, we also use the Tinker \textit{et al.} fitting function with the same parameters values, however we use the linear power spectrum computed in MG to evaluate the relation between the mass and the peak height.} \cite{tinker2010large}, and 
$\xi^L_m(r)$ is the linear matter correlation function calculated from the Fourier transform of the MG linear power spectrum $P^L_m(k)$ obtained from \texttt{MGCAMB} \cite{hojjati2011testing},
\begin{equation}
\xi^L_m(r)=\frac{1}{2\pi^2}\int dk \, k^2 \, P^L_m(k) \frac{\sin (kr)}{kr}.
\label{eq:corrfunc}
\end{equation}
In what follows we will identify the halo mass with the virial mass, $M_h \to M_v$.

\section{Turnaround radii and masses of structures}
\label{sec:4}
In the Spherical Collapse Model (SCM) the formation of gravitational structures takes place shell by shell, with each shell reaching its turnaround and, subsequently, collapse \citep{mo2010galaxy}.
Although in GR each shell evolves independently, due to Birkhoff's theorem, this is not true anymore in the MG framework, since that theorem is no longer valid \cite{Faraoni1}. Hence, the behavior of each shell will be influenced by all others, as shown by, e.g., \citep{lopes2018turnaround,brax2012structure, martino2009spherical, PhysRevD.88.084015, PhysRevD.85.063518}. 

Our goal in this Section is to relate the turnaround mass $M_t$ of the shell which is at the turnaround moment with the virial mass $M_v$ of the shell that, at that same moment, has reached virialization. For that end we employ the criterium that the density inside the inner shell should be $\Delta = 200$ times the density of the background. Using the expression \eqref{eq:turnradius} we write the ratio $M_t/M_v$ as 
\begin{equation}
\frac{M_t}{M_v}=\left(\frac{R_t} {R_v}\right)^3\left(\frac{1+\delta_t(a_t, M_t )}{\Delta}\right).
\label{eq:mastovermasv}
\end{equation}
By considering a top-hat profile, Refs. \cite{cupani2008mass, eke1996cluster} presented an expression for $R_t/R_v$, however in those works the dependence of the collapse not only on radius, but also on the mass of the structures, was not considered.

As discussed above, the turnaround and virial scales can be connected using the NFW density profile and the two-halo term. In this context, the turnaround mass $M_t$, from Eq. \eqref{eq:turnradius}, can be expressed as a function of the virial mass by
\begin{equation}
M_t = 4\pi\int_0^{R_t} dr \, r^{2} \, \left[\rho_{NFW}(r|M_v)+\rho_{2h}(r|M_v)\right] ,
\label{eq:MturnMvir1}
\end{equation}
or, applying Eqs.\eqref{eq:massmax2}-\eqref{eq:2haloterm}, 
\begin{align}
M_t & = 4 \pi \rho_s \left[\frac{R_{v}}{c(z, M_{v})}\right]^3
\left\{\ln\left[1+c(z, M_{v}) \frac{R_t}{R_v} \right]-\frac{c(z, M_{v}) \, \frac{R_t}{R_v}}{ c(z, M_{v}) \, \frac{R_t}{R_v}+1}\right\} \nonumber \\
& + 4\pi \bar{\rho}_{m} \, b^L(M_v) \int_0^{R_t} dr \, r^2 \, \xi^L_m(r) \; .
\label{eq:MturnMvir2}
\end{align}
Considering that the virial mass of the structure can be calculated through \eqref{eq:massmax2}, we can now rewrite the normalization $\rho_s$ in terms of 
the virial mass, such that the turnaround mass is
\begin{align}
M_t & = M_v 
\left\{
\frac{\ln\left[1+c(z, M_{v}) \frac{R_t}{R_v} \right]
-\frac{c(z, M_{v}) \frac{R_t}{R_v}}{c(z, M_{v}) \frac{R_t}{R_v}+1}}
{\ln\left[1+c(z, M_{v}) \right]-\frac{c(z, M_{v})}{c(z, M_{v})+1}}
\right\} + \nonumber \\
& + \frac{2}{\pi} \Omega_{m0} \, \rho_{cr0} \, b^L(M_v) R_t \int_0^{\infty} dk \, \left[\frac{\sin(k R_t)}{k R_t}-\cos(k R_t) \right] P(k) \; .
\label{eq:MturnMvir3}
\end{align}
Combining this last relation with Eq. \eqref{eq:turnradius}, we can solve the two equations to find $R_t/R_v$ for each value of the virial mass $M_v$. 
Additionally, we can also express $M_t$ in terms of $M_v$ with the help of Eq. \eqref{eq:mastovermasv}.

\subsection{Ratio between the Turnaround radius and Virial radius}
\begin{figure}[ht]
\centering
\includegraphics[width=2.5in]{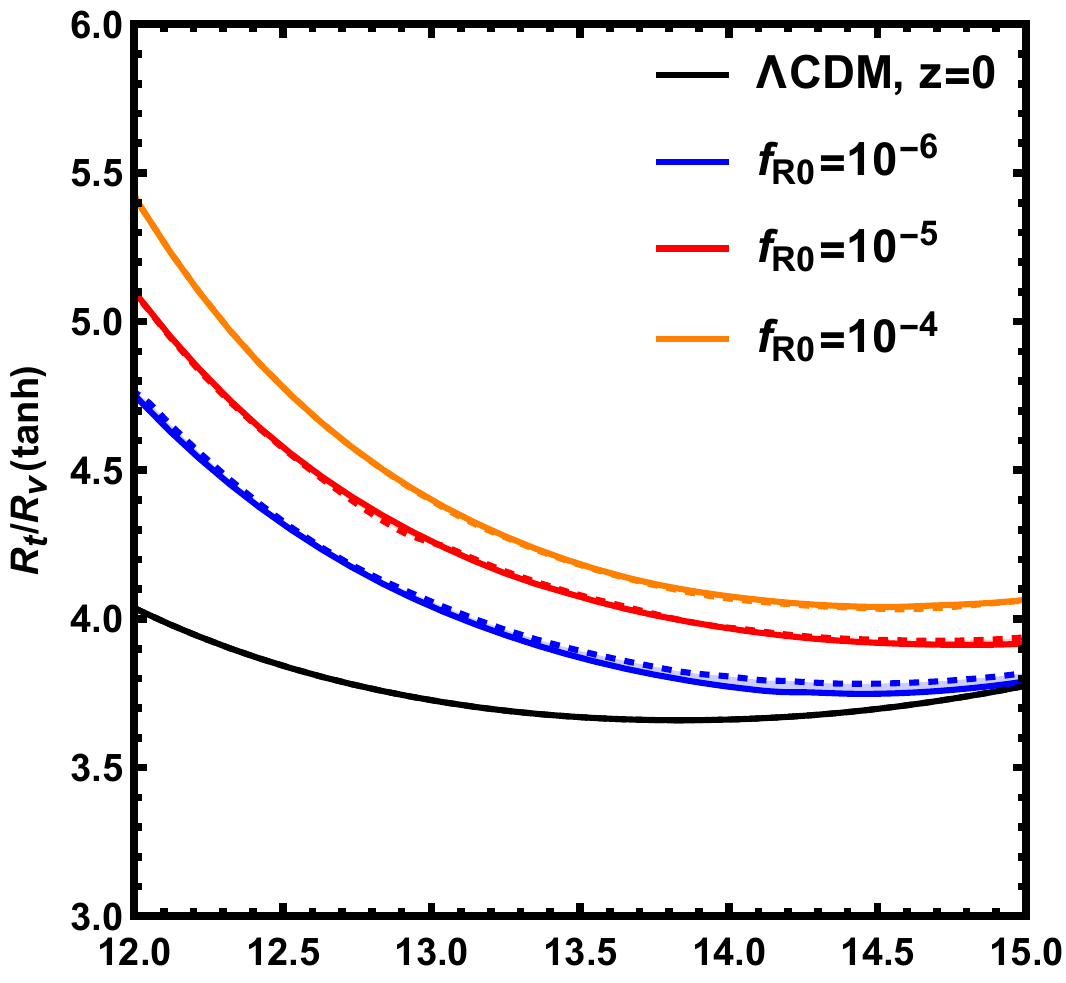}
\includegraphics[width=2.5in]{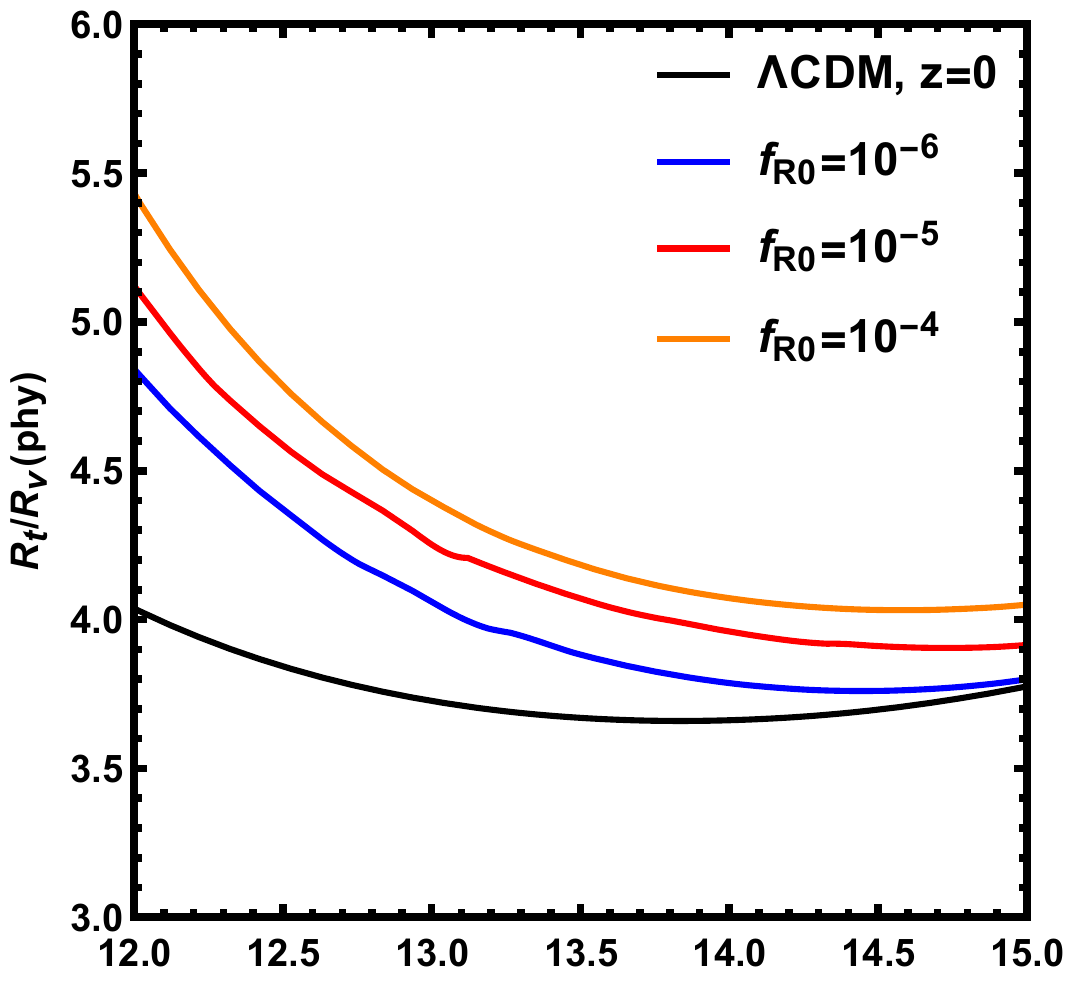}
\includegraphics[width=2.5in]{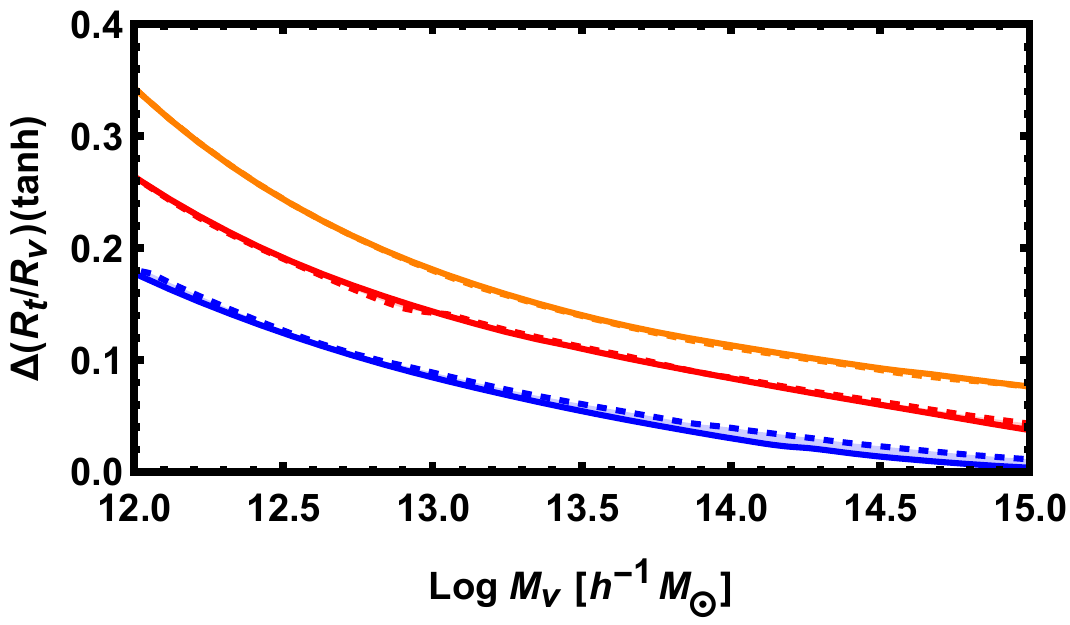}
\includegraphics[width=2.5in]{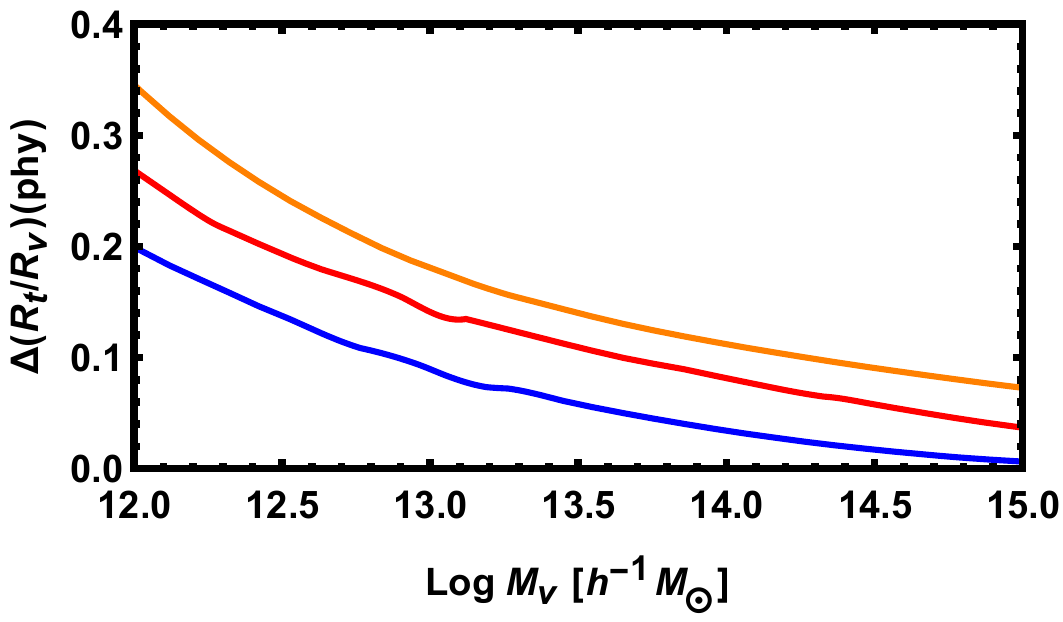}
\caption{\small{Top panels, from left to right: ratios $R_t/R_{v}$ for the Tanh and Phy initial profiles, respectively, in the range $10^{12} \, h^{-1} \, M_\odot<M_{v}<10^{15} \, h^{-1} \, M_{\odot}$, and for MG parameters $|f_{R0}|=10^{-4}$ (orange), $|f_{R0}|=10^{-5}$ (red), $|f_{R0}|=10^{-6}$ (blue), and for $\Lambda$CDM, $\epsilon=0$ (Black). The Tanh profile with a smooth slope ($s=0.8$) is plotted as the dotted lines, and the hard slope ($s=0.4$) is plotted as the solid lines.
Lower panels: relative differences between the values of the top panels with respect to $\Lambda$CDM.}}
\label{fig:ratioRtRvMv}
\end{figure}

As previously stated, in our approach we have considered the overdensity $\Delta=200$, widely used in observational practice. 
Fig. \ref{fig:ratioRtRvMv} shows how $R_t/R_v$ depends on the mass $M_v$  for the Tanh (left panel) and the Physical (right panel) initial profiles, for masses ranging from $10^{12}h^{-1}M_\odot$ (small groups of galaxies) to $10^{15}h^{-1}M_\odot$ (massive clusters), which includes most cosmic structures of interest. 
We focus on the mass $10^{13}h^{-1}M_\odot$ (galaxy groups) because it seems well suited to study the effects on the turnaround radius \cite{Tanoglidis:2014lea}: in fact, tests involving the turnaround are more sensitive around small structures, where the effects of MG are enhanced. 
On the other hand, it is also important to consider structures whose turnarounds can be estimated with good accuracy -- which is again the case for small galaxy groups \cite{Tanoglidis:2014lea}.

For the plots of Fig. \ref{fig:ratioRtRvMv} we used values of the modified gravity parameter $f_{R0}=10^{-6}$ (blue), $f_{R0}=10^{-5}$ (red), $f_{R0}=10^{-4}$ (orange), as well the $\Lambda$CDM model ($f_{R0} \to 0$, or $\epsilon \to 0$ -- black).
In the case of the Tanh initial profile (left panels) we plot two slopes, $s=0.4$ (solid lines) and $s=0.8$ (dotted lines),
but the behavior in both cases is similar, and very close to that of the Phy initial profile (right panels). 
The ratio $R_t/R_v$ increases for smaller masses, and is larger in MG compared with $\Lambda$CDM -- for our reference mass scale
$M=10^{13}h^{-1}M_{\odot}$, we have $R_t=4.1 R_{v}$ for $f_{R0}=10^{-6}$, and $R_t=3.7 R_{v}$ for $\Lambda$CDM.

Interestingly, we obtained values of $R_t/R_{v}$ which agree with the measurements of Ref. \cite{kashibadze2018cosmic}. 
There the authors estimated the radius of the zero-velocity surface, $R_t$, through the data of distances and radial velocities of galaxies on the Local Group, and for a stacking of $14$ nearby groups, showing that $R_t \approx (3-4) R_{v}$. However, we need more accurate observations if we wish to distinguish $\Lambda$CDM from the MG models considered in our work.

On the bottom of  Fig. \ref{fig:ratioRtRvMv} we have plotted the relative differences of the ratio of the radii as a function of the virial mass. Even for a small deviation from GR, $f_{R0}=10^{-6}$, the relative difference of the two for our reference mass scale $M=10^{13} \, h^{-1} \, M_{\odot}$ is $\sim 10\%$, while for $f_{R0}=10^{-5}$ the difference is $\sim 18\%$. 
Hence, if it becomes possible to measure the turnaround radius and the virial radius independently, with precision of $\simeq 10\%$ in their ratio, then we could use this to constrain this class of MG models down to $f_{R0} = 10^{-6}$, which is a very competitive result \citep{burrage2018tests}.  


\subsection{Ratio between the Turnaround mass and Virial mass}

\begin{figure}[ht]
\centering
\includegraphics[width=2.5in]{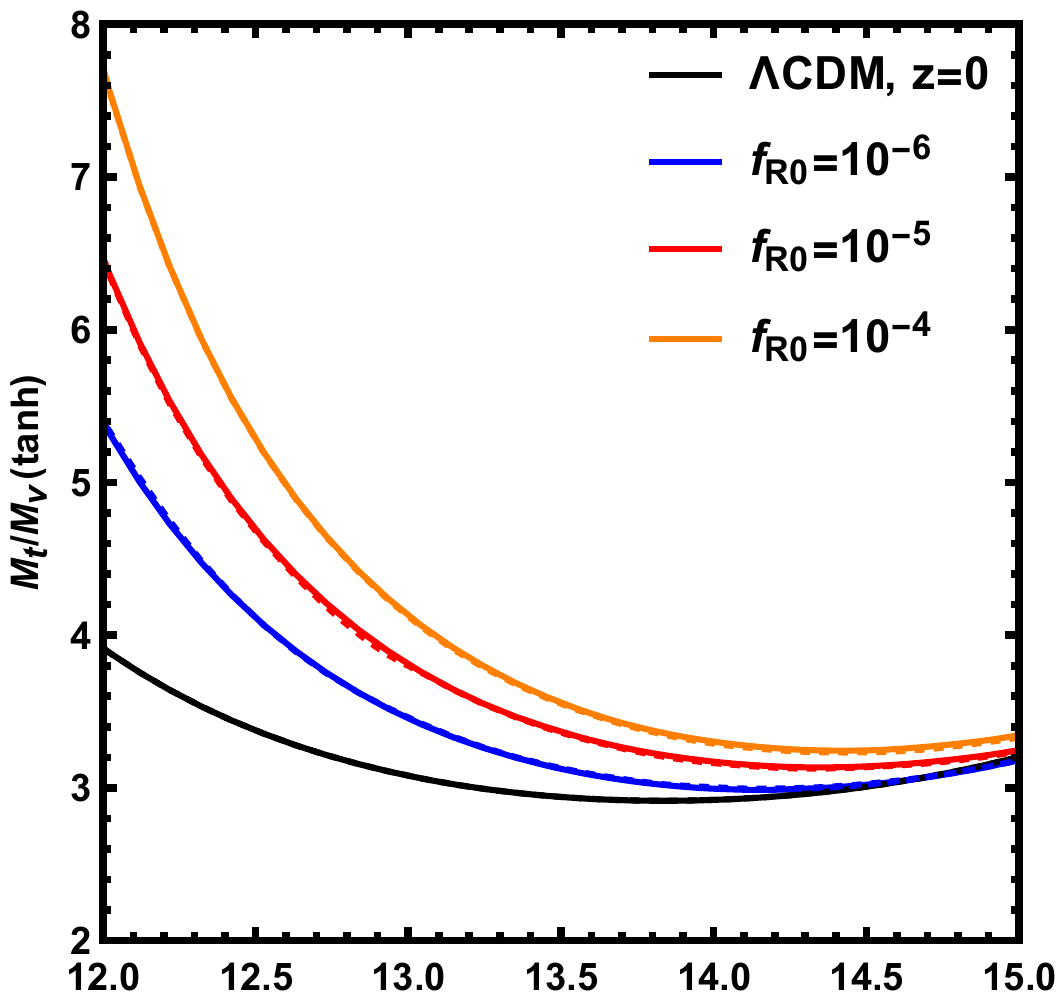}
\includegraphics[width=2.5in]{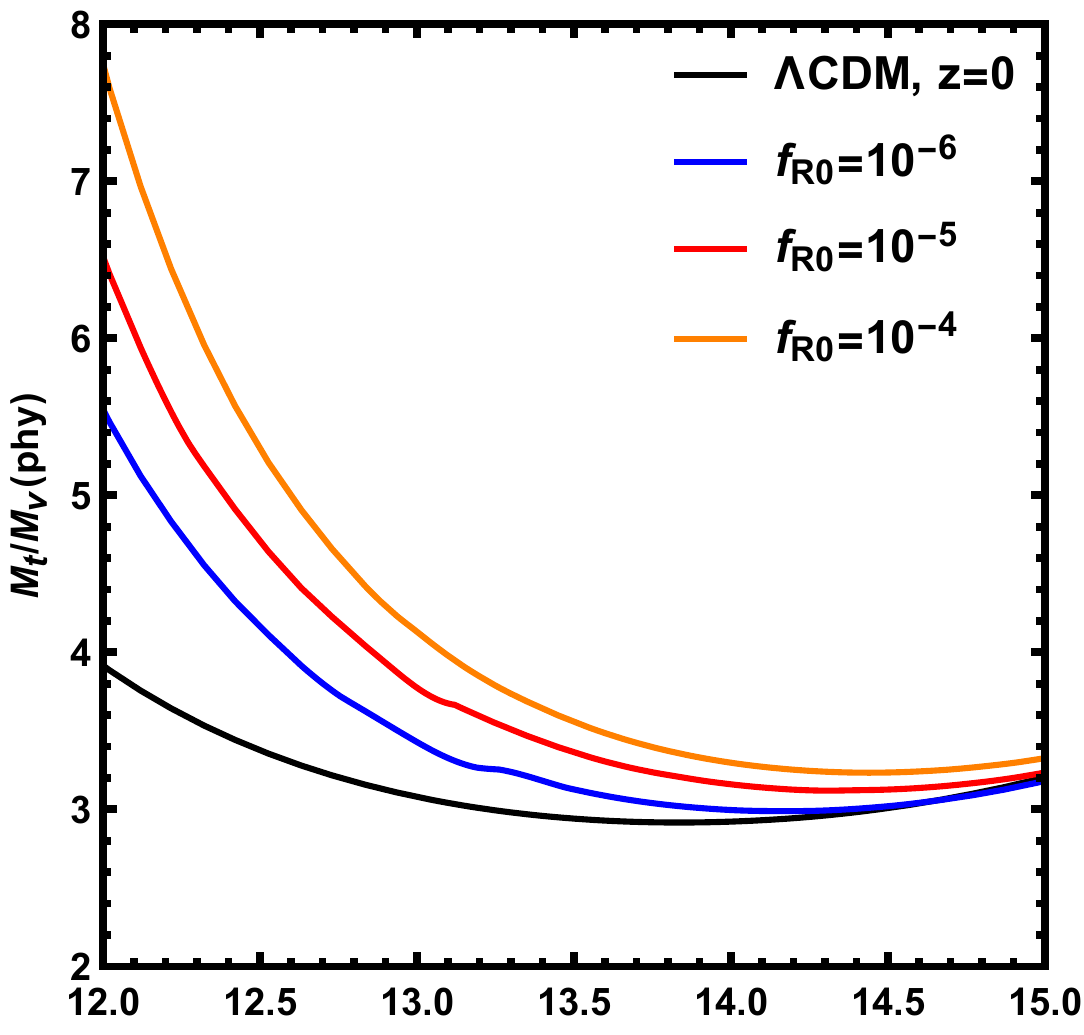}
\includegraphics[width=2.5in]{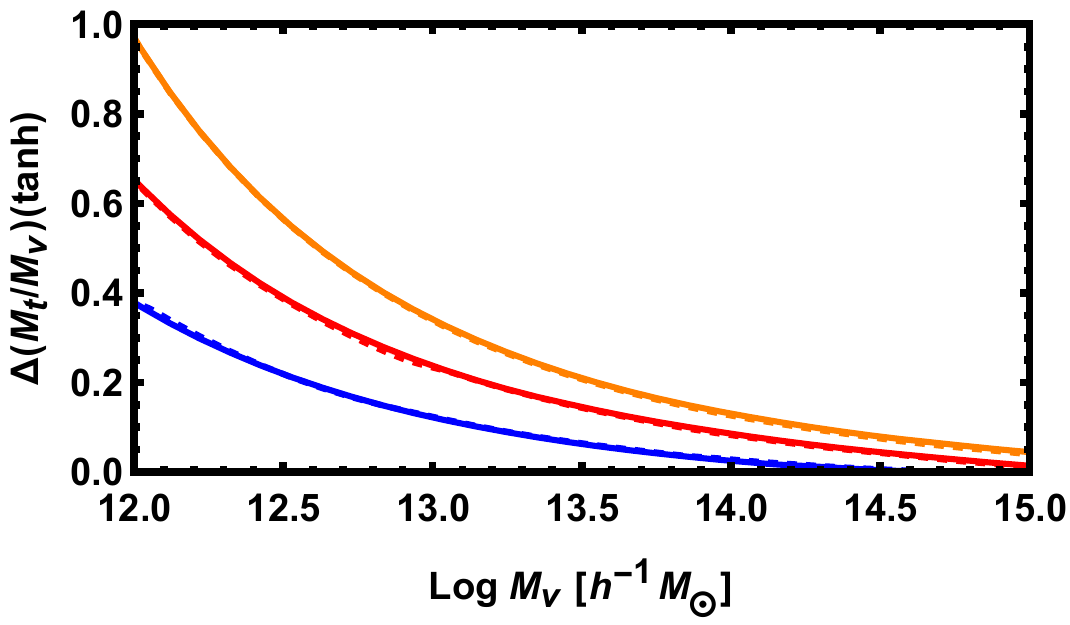}
\includegraphics[width=2.5in]{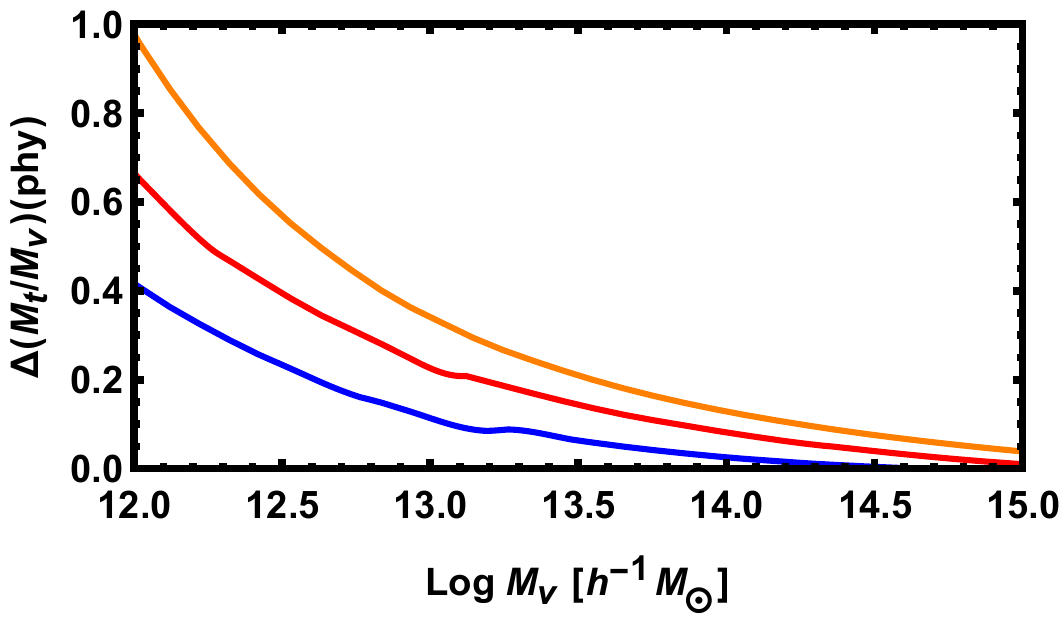}
\caption{\small{Top panels, from left to right: $M_t/M_{v}$ for the Tanh and Phy initial profiles, respectively, for $10^{12} \,  h^{-1} \, M_\odot<M_{v}<10^{15} \, h^{-1} \, M_{\odot}$, and for MG parameters $|f_{R0}|=10^{-4}$ (orange), $|f_{R0}|=10^{-5}$ (red), $|f_{R0}|=10^{-6}$ (blue), as well as $\Lambda$CDM or $\epsilon=0$ (Black). As in previous plots, the Tanh profile with a smooth slope ($s=0.8$) is plotted as the dotted lines, and the hard slope ($s=0.4$) is plotted as the solid lines.
Lower panels: relative differences between the values of the top panels with respect to $\Lambda$CDM.}}
\label{fig:ratioMtMvMv}
\end{figure}

Just as was done for the turnaround and virial radii, it is also possible to obtain a relationship between the turnaround mass and the virial mass using Eq. \eqref{eq:mastovermasv}. 
The results of this process are presented in Fig. \ref{fig:ratioMtMvMv}, where we can see in which way $M_t/M_{v}$ depends on the virial mass of the gravitational structures. 
Within the range of masses we considered, this ratio reaches a minimum at the mass scale of $M \simeq 10^{13.5}  \, h^{-1}  \, M_{\odot}$ for $\Lambda$CDM ($M_t/M_{v} \simeq 2.9$), and at the mass $M \simeq 10^{13.8} \, h^{-1} \, M_{\odot}$ for $|f_{R0}|=10^{-6}$ ($M_t/M_v \simeq 3.0$). 
We also note $M_t/M_{v}$ and $R_t/R_{v}$ are steeper functions of virial mass at the low end, compared with standard cosmology. 
For structures of mass $M=10^{13} \, h^{-1}  \, M_{\odot}$ the relation is $M_t=3.07 M_{v}$ for $\Lambda$CDM, while for $|f_{R0}|=10^{-6}$  it is $M_t=3.43 M_{v}$.
As for the difference with respect to $\Lambda$CDM, for our reference mass scale it reaches $\simeq 11\%$, $\simeq 22\%$ and $\simeq 34\%$ for $|f_{R0}|=10^{-6}$, $|f_{R0}|=10^{-5}$ and $|f_{R0}|=10^{-4}$, respectively. 

\section{Turnaround radius for realistic structures in \texorpdfstring{$f(R)$}{f(R)}}
\label{sec:5}

\subsection{Turnaround and virial mass: theory}
\label{sec:5.1}

\begin{figure}[ht]
\centering
\includegraphics[width=2.5in]{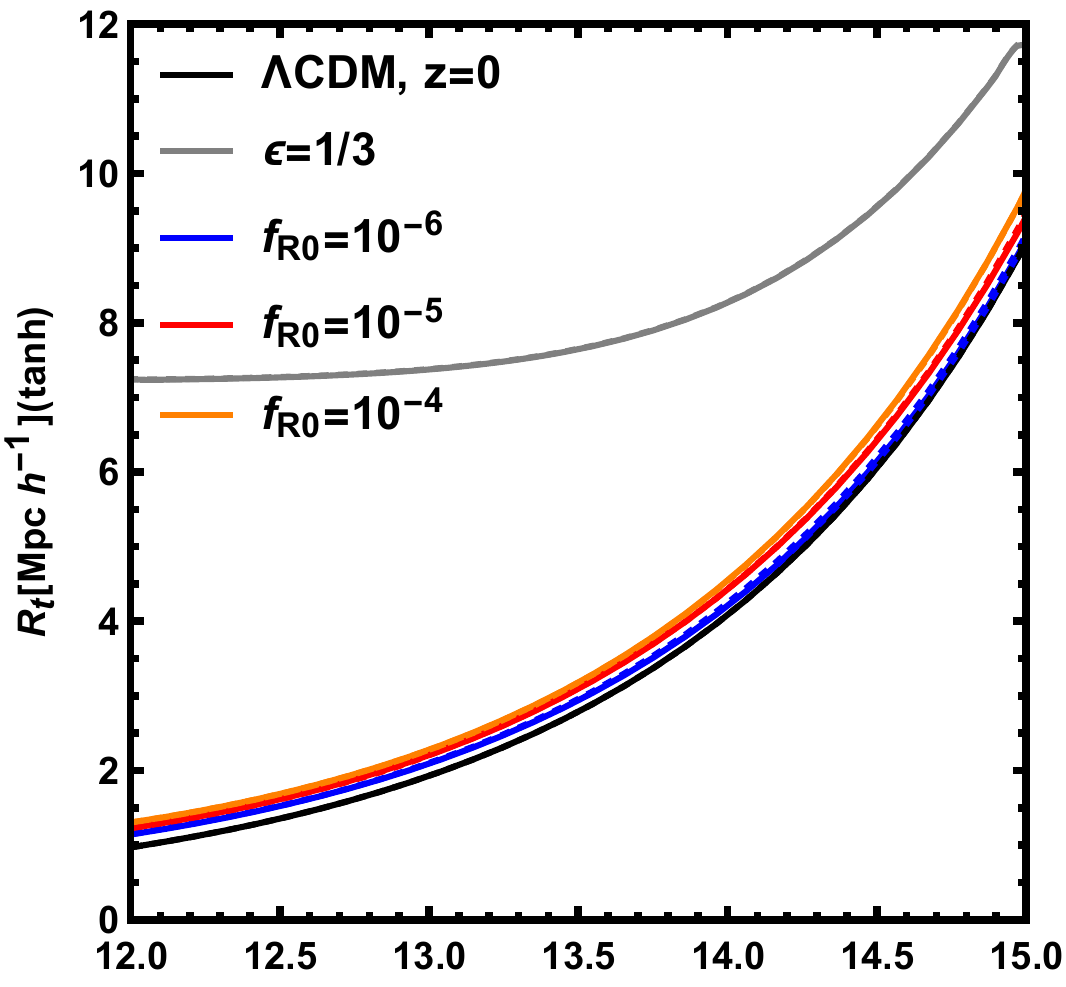}
\includegraphics[width=2.5in]{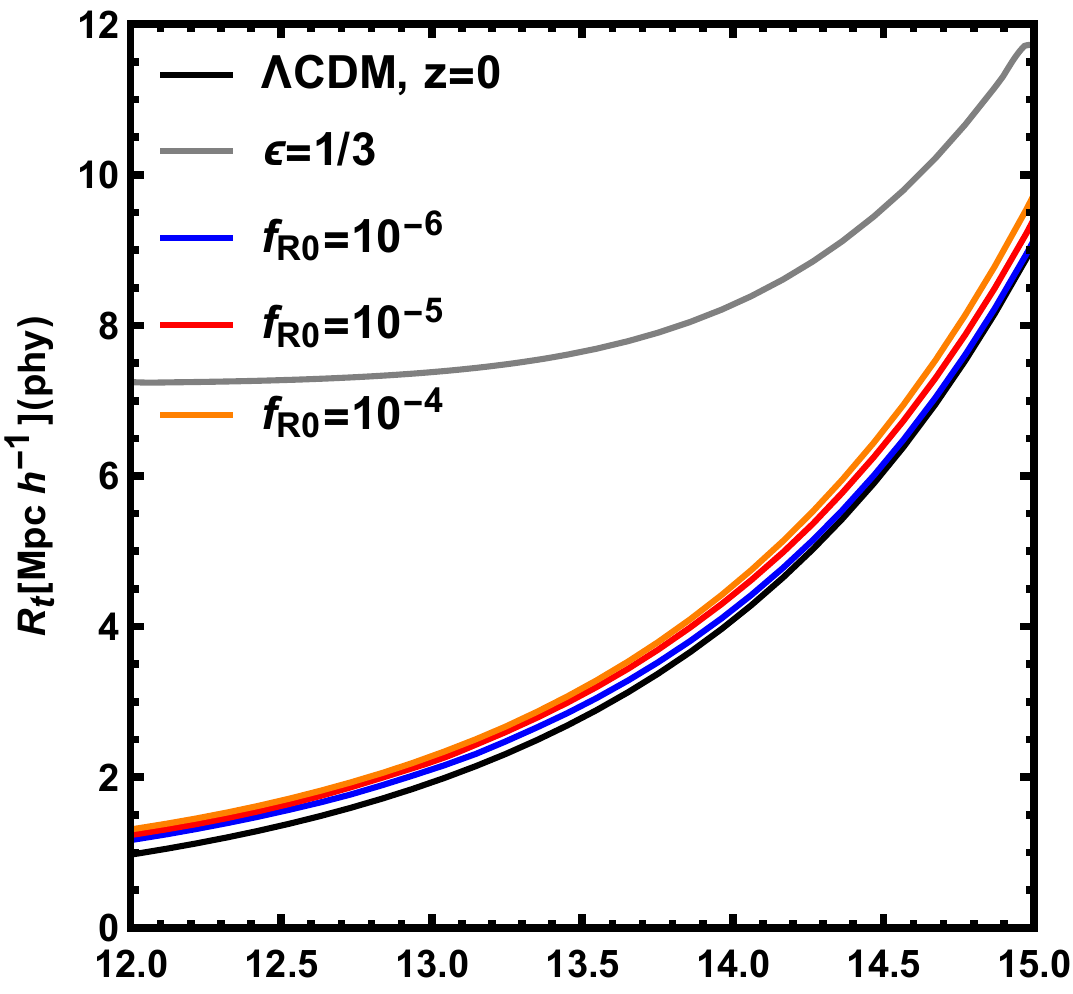}
\includegraphics[width=2.5in]{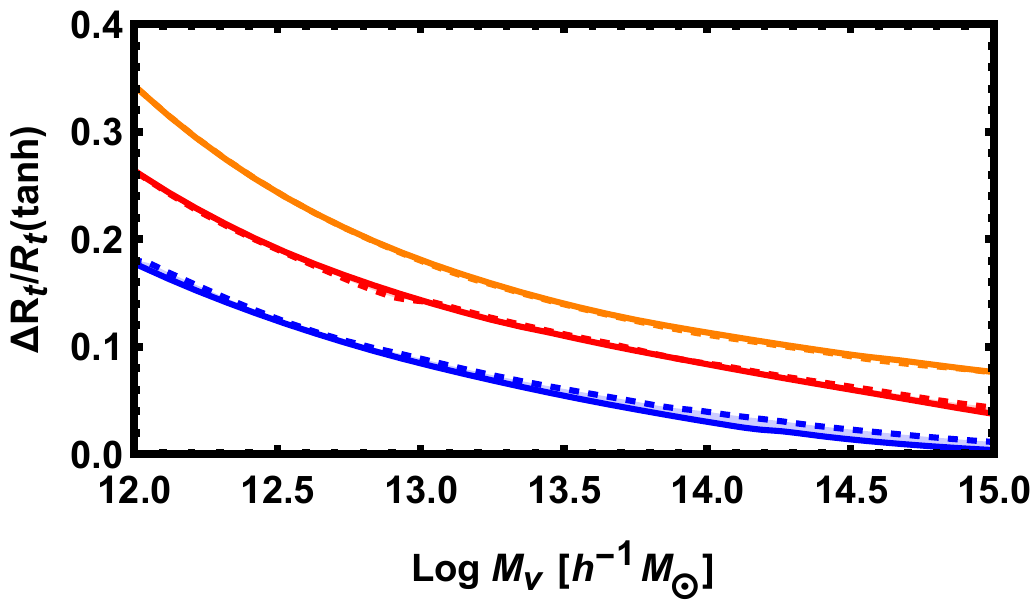}
\includegraphics[width=2.5in]{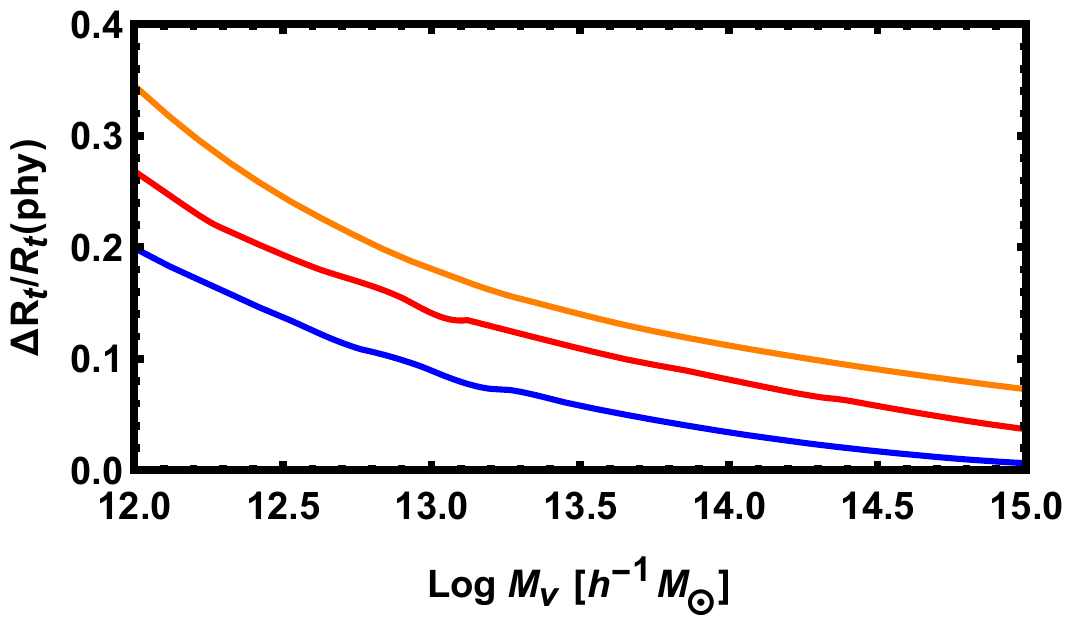}
\caption{\small{Top panels, from left to right: values of $R_t$ for the Tanh and Phy profiles, respectively, for $10^{12}h^{-1}M_\odot<M_{v}<10^{15}h^{-1}M_{\odot}$, and for MG parameters $|f_{R0}|=10^{-4}$ (orange), $|f_{R0}|=10^{-5}$ (red), $|f_{R0}|=10^{-6}$ (blue) $\Lambda$CDM, $\epsilon=0$ (Black). The Tanh profile with a smooth slope ($s=0.8$) is plotted as the dotted lines, and the hard slope ($s=0.4$) is plotted as the solid lines.
Lower panels: relative differences between the values of the top panels with respect to $\Lambda$CDM.}}
\label{fig:compadcZdifRel}
\end{figure}

Clearly, the most convenient variables, from an observational perspective, are the turnaround radius and the virial mass, and this
relationship can be computed with the help of Eq.\eqref{eq:turnradius}.
The dependence of the turnaround radius with virial mass is the central result of this paper, and is plotted in Fig. \ref{fig:compadcZdifRel} for $\Lambda$CDM as well as for the three MG models that we have analyzed in this paper. 
That figure also presents the results in the large-field limit, which was omitted in previous plots for the sake of brevity. 
The large differences with respect to $\Lambda$CDM (mainly for the large-field limit) are generated because of the additional difference between the density profiles (Fig. \ref{fig:profiles}). 
In fact, when we focus on the spherical collapse parameter $\delta _{t}$, this difference is not larger than $\simeq 20 \%$ (see figures 5 and 6 of \cite{lopes2018turnaround}).

These results can be used for testing MG with current measurements, and they can be checked with simulations. Indeed, from the lower panels in  Fig. \ref{fig:ratioMtMvMv}, which show the relative differences with respect to $\Lambda$CDM, we notice that at our reference mass scale of $10^{13} \, h^{-1} \, M_{\odot}$, the turnaround radius is larger by $\simeq 9 \%$, $\simeq 14 \%$ and $\simeq 18 \%$, for $|f_{R0}|=10^{-6}$, $|f_{R0}|=10^{-5}$ and $|f_{R0}|=10^{-4}$, respectively. 

\subsection{Turnaround and virial mass: comparison with data}
\label{sec:5.2}

The link between the observable turnaround radius (i.e., the one we defined here, in terms of the null velocity surface) and the virial mass of collapsed structures which was derived in the previous section allows us to compare our results with data. We have collected different sets of observations of cosmic flows in the nearby Universe from Rines \& Diaferio 2006 \cite{RinesDiaferio}, Pavlidou \& Tomaras 2014 \cite{Pavlidou:2014aia}, and
Lee 2017 \cite{2017arXiv170906903L}. 

\begin{figure}[ht]
\centering
\includegraphics[height=4.0in]{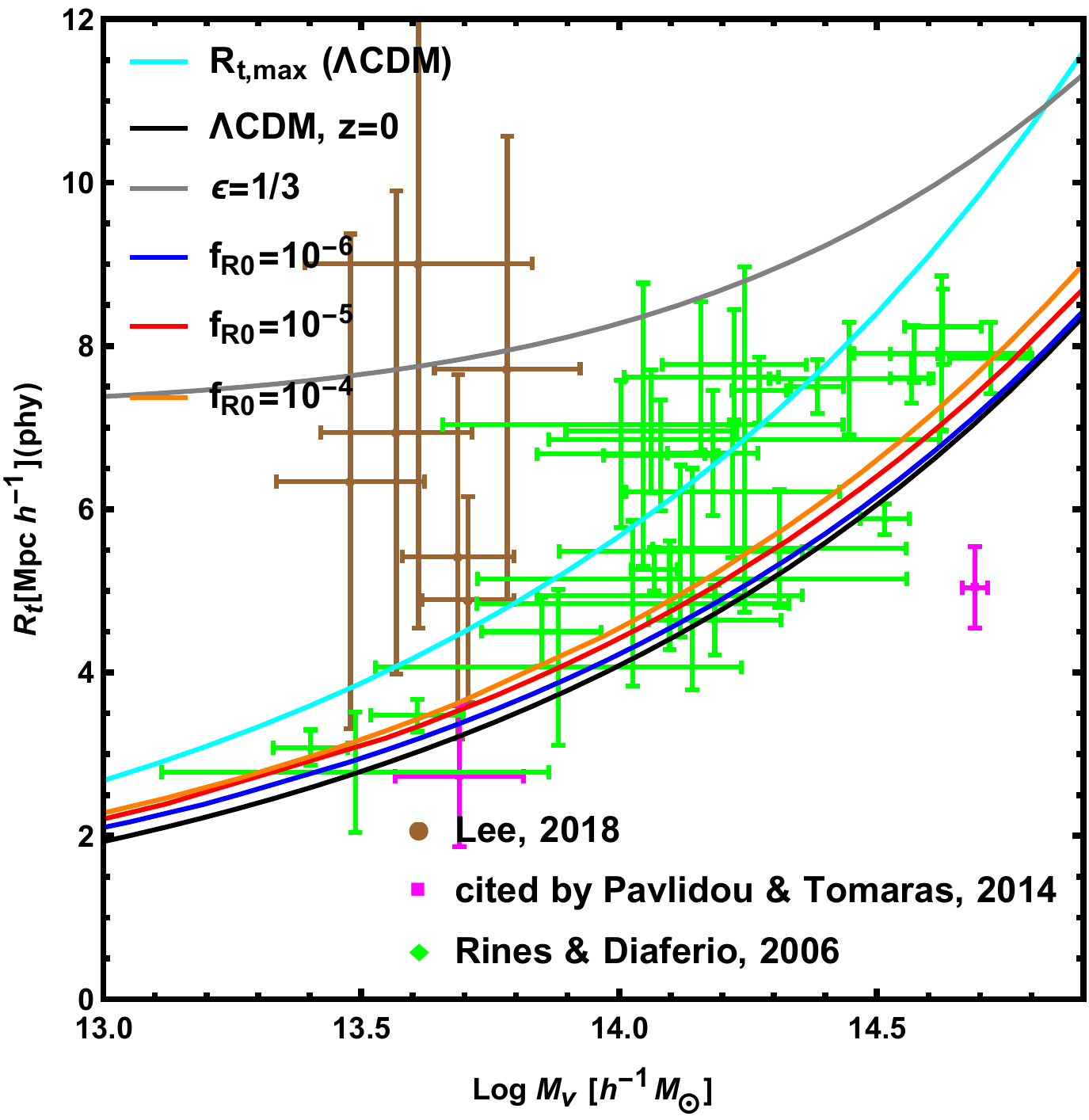}
\caption{\small{Theoretical predictions for the turnaround radius for $\Lambda$CDM (black solid line) and MG (colored solid lines), including the ``maximum turnaround radius'' (cyan solid line; see text). Data points taken from Refs. \citep{RinesDiaferio,Pavlidou:2014aia,2017arXiv170906903L}.}}
\label{fig:theoryvdata}
\end{figure}

In Fig. \ref{fig:theoryvdata} we present the comparison between theory (solid lines) and data around the mass scales corresponding to these systems. As usual, we show the theoretical expectations for GR and for the Hu-Sawicky model with several strengths of the MG parameter $f_{R0}$ -- including the extreme case, $\epsilon = 1/3$. For completeness we also show the so-called ``maximum turnaround radius'', a bound that can be derived analytically for the $\Lambda$CDM model \citep{Bhattacharya2017cosmic, lee2017effect, bhattacharya2017maximum}.

In particular, Lee 2017 \cite{2017arXiv170906903L} has measured cosmic flows around groups in the local universe with masses $[0.3 - 1] \times 10^{14} M_{\odot}$, reporting uncertainties in the turnaround radius of order  $25 - 50 \%$. 
Those measurements appear inconsistent with the turnaround radius in $\Lambda$CDM or any of the MG models considered here -- beside violating the maximum bound of Refs. \citep{Bhattacharya2017cosmic, lee2017effect, bhattacharya2017maximum}.
Even as Ref. \cite{2017arXiv170906903L} claims to measure turnaround radii which are larger compared to the models, the systems examined by Pavlidou \& Tomaras 2014 \cite{Pavlidou:2014aia} seem to indicate turnaround radii substantially below theoretical predictions.
Rines and Diaferio 2006 \cite{RinesDiaferio}, on the other hand, estimate the turnaround radius from a large number of groups and clusters using both a theoretical fit and an estimate of the ``maximal'' value of the radius of the caustic region corresponding to null peculiar velocities (and for this reason we quote as uncertainties the interval between the theoretical fit and the maximal value of the turnaround).

From Fig. \ref{fig:theoryvdata} it appears that there are systematic issues with measurements of the turnaround radius, and for this reason we have not binned the data points.
Here we must pause and ponder that both theory and data, at this level, cannot be over-interpreted. First, from the point of view of the simulations, realistic systems are not spherically symmetric (a key assumption underlying our results), nor typical initial profiles are necessarily well represented by our Phy and Tanh density profiles. This issue, however, {\em can} be addressed, e.g., by employing N-body simulations in $\Lambda$CDM and in MG (Voivodic et al., in preparation). Second, from the point of view of observations, often there are many systematic issues which can only be identified and controlled with more accurate and precise measurements. These measurements should aim at a wide range of mass scales, and ideally they should focus on structures which are as isolated as possible. If, with these improvements, we are able to reach uncertainties at the level of $\sim$ few percent for stacks of systems in a few bins of virial mass between $10^{13}-10^{15} \, h^{-1} \, M_\odot$, we will be in a position of imposing robust and competitive constraints on MG using the turnaround radius.

Another important issue, which can be addressed with the help of N-body simulations, is the relation between the turnaround and the splashback radius. The splashback radius determines the transition from the one-halo to the two-halo term \cite{Snaith}, while the turnaround radius expresses the scale where the halo decouples from the background
-- see, e.g., the measurements performed by Ref. \cite{Chang}. In the literature is not clear what is the most suitable radius to characterize the halos, and the splashback radius is also a good candidate \cite{More}. Other works also show that the splashback radius has a counterpart in the velocity field and, therefore, in the turnaround radius \cite{Okumura}, and that this scale changes with different gravitational and dark energy theories \cite{Adhikari}. Therefore, the relationship between the two radii can provide complementary information, since the turnaround radius is measured through the velocity field, while the splashback radius is determined by the density profile.

\section{Conclusions}
\label{sec:conc}
In order to compare the results obtained in \cite{lopes2018turnaround} with future measurements in simulations and observations, we developed a new approach to relate a structure which is near the virialization with an outer shell that is in the turnaround moment.
As observed by \cite{kashibadze2018cosmic}, the number of galaxies between $R_v$ and $R_t$ is approximately $\sim 15\%$ of the number of galaxy inside $R_t$, therefore the mass of the structure in the virial moment is much easier to measure than the turnaround mass, and a relation between $R_t$ and $M_v$ can be a useful tool to test models of modified gravity. 
We should note that an expression that links $R_t$ and $R_v$ was also obtained by \cite{eke1996cluster}; however that result was obtained by implementing the spherical collapse model without considering the mass profile inside the surface of null radial velocity, as has been done here. In our case, we use a specific profile, the NFW profile plus the 2-halo term, which makes our result more realistic.

Our findings show that for structures with mass of $10^{13} \, h^{-1}\, M_\odot$, the turnaround radius and virial mass are related to the same virial quantities by $R_t=3.7 \, R_{v}$  and $M_t=3.07 \, M_{v}$ for the standard $\Lambda$CDM cosmology, while for modified gravity, according to the prescription of Hu \& Sawicki, parametrized by $|f_{R0}|=10^{-6}$, these relations become  $R_t=4.1 \,  R_{v}$  and $M_t=3.43 \,  M_{v}$ -- i.e., the relative difference of MG with respect to the $\Lambda$CDM model is $\sim 10\%$ for the radii and $\sim 11\%$ for the masses. Therefore, if we have measurements of these quantities with accuracy of $\lesssim 10\%$, it should be possible to use them to test modified gravity. 

Moreover, as the main result of this work, we compute the direct relation between the turnaround radius $R_{t}$ and the halo mass $M_{v}$ for the Hu \& Sawicki model of MG. Note that, different from the relation between the radii and masses, the turnaround radius and, of course, the halo mass are already observables, and can be directly used to constraint MG. We found an enhancement of $9\%$ in the turnaround radius, for a virial mass of $10^{13} \, h^{-1}\, M_{\odot}$, in the context of the weakest version of MG considered here ($f_{R0} = 10^{-6}$). Current observations are reaching the uncertainty levels necessary to put constraints in MG parameters, but in order for the turnaround radius to become a robust and competitive test, both theoretical predictions need to become more realistic, and systematic differences between different measurements need to be addressed.

As future work we want to test our predictions using N-body simulations, where all observables are accessible. With an N-body simulation it is possible to test the predictions made here and those obtained in Ref. \cite{lopes2018turnaround}, since we will have access to the virial and turnaround radii, as well as the turnaround and virial masses of each halo. Moreover, with the simulations it is possible to predict the accuracy of observations by populating the dark matter halos with galaxies. With that knowledge we will be able to determine the observational needs (such as the survey's area, depth, and redshift accuracies) so that the resulting measurements can lead to robust constraints on MG models. Finally, we want to use the data from future spectroscopic surveys like DESI \cite{DESI}, as well as multi-band photometric surveys like JPAS \cite{JPAS}, to measure the velocity fields around clusters.

\appendix
\section{Density profile}
\label{append}
\begin{figure}[ht]
\centering
\includegraphics[width=3.5in]{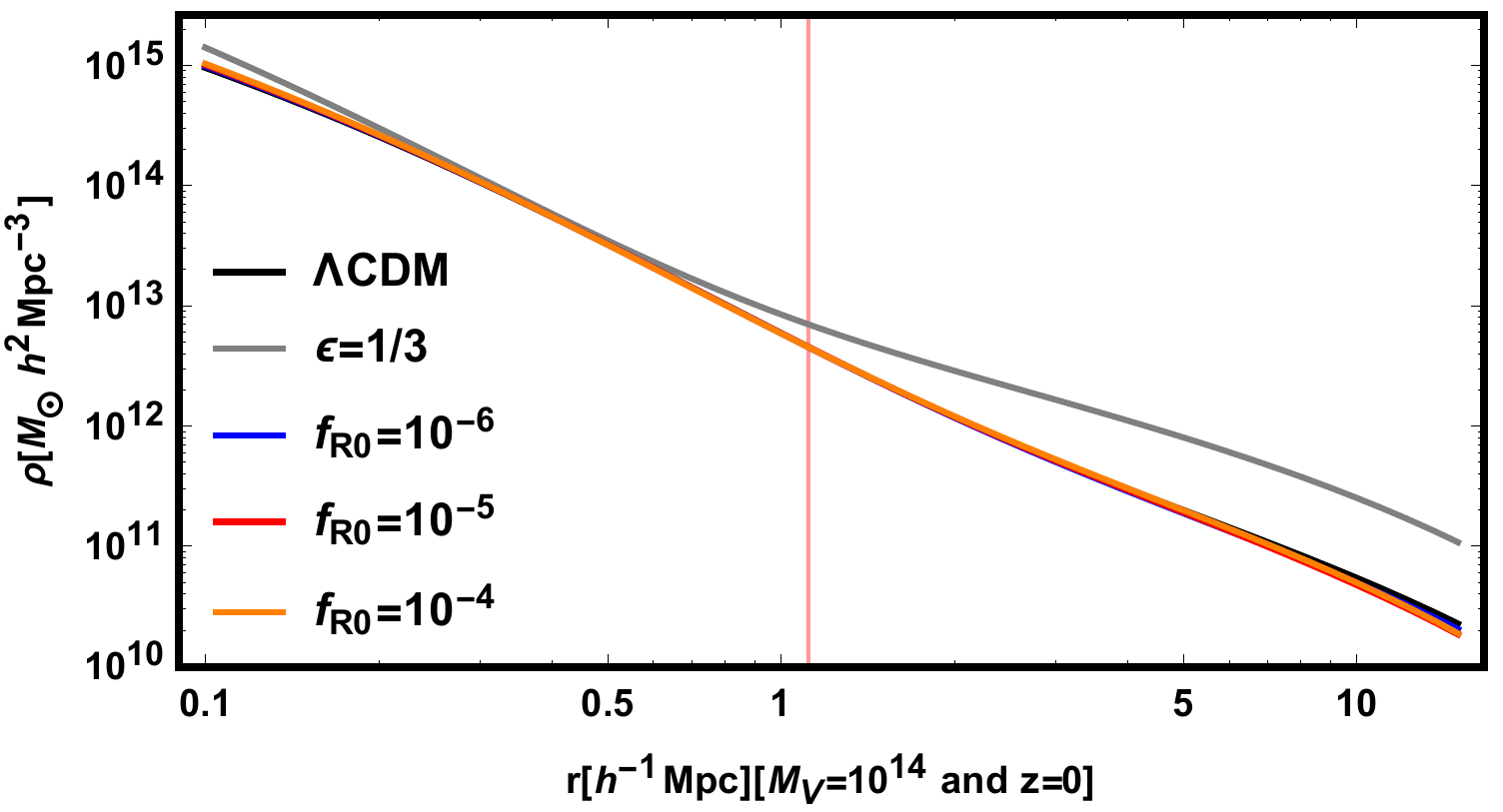}
\includegraphics[width=3.5in]{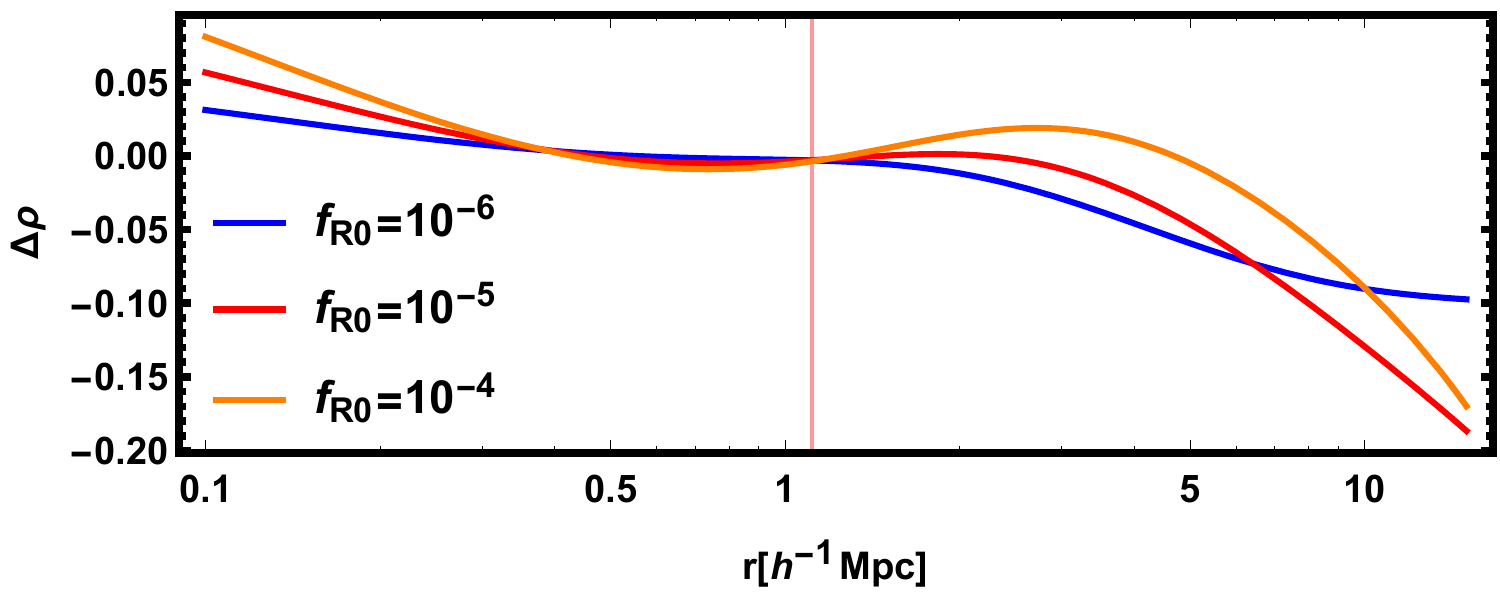}
\caption{\small{Top panel: The density profile composed by the NFW profile and the 2-halo term for $M_v=10^{14}h^{-1}M_\odot$ and $z=0$, considering the MG parameters $|f_{R0}|=10^{-4}$ (orange), $|f_{R0}|=10^{-5}$ (red), $|f_{R0}|=10^{-6}$ (blue), the small-field limit $\epsilon=0$ ($\Lambda$CDM, black) and the large-field limit $\epsilon=1/3$ (gray). The vertical orange line corresponds to $R_v$.
Lower panel: relative differences between the values of the top panel with respect to $\Lambda$CDM.}}
\label{fig:profiles}
\end{figure}

In Fig. \ref{fig:profiles} we present the density profile \eqref{eq:density} for a halo with mass $10^{14} h^{-1}M_{\odot}$ at $z = 0$. We show the profiles for the five different MG parameters used in this paper, including the small and large field limits. On the bottom panel is the relative difference of the profiles with respect to the $\Lambda$CDM case.

There are two effects of MG in the density profiles: the change in $M_{\star}$ of the concentration relation of Bullock \textit{et al.} \cite{bullock2001profiles}, and the change in the two halo term because of the different linear matter power spectrum.

The change in $M_{\star}$ occurs because, in MG, the collapse parameter $\delta _{c}$ depends on the halo mass, however, the variance of the linear field $S(M) = \sigma ^{2} (M)$ will also change due to the modified linear matter power spectrum. These changes will increase $M_{\star}$ as the strength of modified gravity grows, which in turn implies a more concentrated halo. This effect can be seen in Fig. \ref{fig:profiles}, where the inner parts of the halo profiles differ by $\approx 10 \%$.

The change in the linear matter power spectrum also affects the 2-halo term, where the new $\sigma (R)$ relation gives a new linear halo bias (even using the same parameters of $\Lambda$CDM) and directly gives a new two-point correlation function. The linear bias will be larger for stronger modifications of gravity, while the correlation function will be smaller, on large scales, to compensate the more concentrated matter on small scales, implying that the net effect of MG does not vary monotonically with $r$. These differences can be seen in the outer parts of the halo in Fig. \ref{fig:profiles}, where $r > R_{v}$ (vertical orange line), with a difference of $\approx 20 \%$ for large radii.

Through inspection of Fig. \ref{fig:profiles} it is clear that the density profiles in MG have a small, but not negligible, difference with respect to the $\Lambda$CDM profiles, and that the modified profiles should be taken into account in order to compute the mass inside some given radius. Moreover, with the results of this works, it is clear that we need to consider the correct density profile for our observables that, in principle, could depends of the tracer, but a new profile is easy to implement in the approach developed in this work.

\acknowledgments
The authors would like to thank Marcos Lima for insightful comments on structure formation in modified gravity models.
This work was supported by the {\it Programa Proqualis} of Instituto Federal de Educa\c{c}\~{a}o, Ci\^{e}ncia e Tecnologia do Maranh\~{a}o (RCCL) and by FAPESP (RV).
LRA and LSJ acknowledge support from FAPESP and CNPq.



%
%
%
%
%
%
%
%
\bibliographystyle{JHEP}
\bibliography{referencias}

\providecommand{\href}[2]{#2}\begingroup\raggedright\begin{thebibliography}{10}

\bibitem{riess1998}
A.~G. Riess et~al., \emph{Observational evidence from supernovae for an
  accelerating universe and a cosmological constant}, {\emph{The Astronomical
  Journal} {\bf 116} (1998) 1009}.

\bibitem{perlmutter1999}
e.~a. Perlmutter, \emph{Measurements of $\omega$ and $\lambda$ from 42
  high-redshift supernovae}, {\emph{The Astrophysical Journal} {\bf 517} (1999)
  565}.

\bibitem{amendola2010dark}
L.~Amendola and S.~Tsujikawa, \emph{Dark energy: theory and observations}.
\newblock Cambridge University Press, 2010.

\bibitem{weinberg1989cosmological}
S.~Weinberg, \emph{The cosmological constant problem}, {\emph{Reviews of Modern
  Physics} {\bf 61} (1989) 1}.

\bibitem{carroll2004cosmic}
S.~M. Carroll, V.~Duvvuri, M.~Trodden and M.~S. Turner, \emph{Is cosmic
  speed-up due to new gravitational physics?}, {\emph{Physical Review D} {\bf
  70} (2004) 043528}.

\bibitem{sawicki2007stability}
I.~Sawicki and W.~Hu, \emph{Stability of cosmological solutions in f (r) models
  of gravity}, {\emph{Physical Review D} {\bf 75} (2007) 127502}.

\bibitem{hu2007models}
W.~Hu and I.~Sawicki, \emph{Models of f (r) cosmic acceleration that evade
  solar system tests}, {\emph{Physical Review D} {\bf 76} (2007) 064004}.

\bibitem{voivodic2016modelling}
R.~Voivodic, M.~Lima, C.~Llinares and D.~F. Mota, \emph{{Modelling Void
  Abundance in Modified Gravity}},
  \href{http://dx.doi.org/10.1103/PhysRevD.95.024018}{\emph{Phys. Rev.} {\bf
  D95} (2017) 024018}, [\href{http://arxiv.org/abs/1609.02544}{{\tt
  1609.02544}}].

\bibitem{PhysRevD.85.063518}
A.~Borisov, B.~Jain and P.~Zhang, \emph{Spherical collapse in $f(r)$ gravity},
  \href{http://dx.doi.org/10.1103/PhysRevD.85.063518}{\emph{Phys. Rev. D} {\bf
  85} (Mar, 2012) 063518}.

\bibitem{PhysRevD.88.084015}
M.~Kopp, S.~A. Appleby, I.~Achitouv and J.~Weller, \emph{Spherical collapse and
  halo mass function in $f(r)$ theories},
  \href{http://dx.doi.org/10.1103/PhysRevD.88.084015}{\emph{Phys. Rev. D} {\bf
  88} (Oct, 2013) 084015}.

\bibitem{Chakrabarti2014}
S.~Chakrabarti and N.~Banerjee, \emph{Spherical collapse in vacuum f(r)
  gravity},
  \href{http://dx.doi.org/10.1007/s10509-014-2118-1}{\emph{Astrophysics and
  Space Science} {\bf 354} (2014) 571--576}.

\bibitem{cembranos2012gravitational}
J.~Cembranos, A.~de~la Cruz-Dombriz and B.~M. Nunez, \emph{Gravitational
  collapse in f (r) theories}, {\emph{Journal of cosmology and astroparticle
  physics} {\bf 2012} (2012) 021}.

\bibitem{herrera2017calculation}
D.~Herrera, I.~Waga and S.~Jor{\'a}s, \emph{Calculation of the critical
  overdensity in the spherical-collapse approximation}, {\emph{Physical Review
  D} {\bf 95} (2017) 064029}.

\bibitem{brax2012structure}
P.~Brax and P.~Valageas, \emph{Structure formation in modified gravity
  scenarios}, {\emph{Physical Review D} {\bf 86} (2012) 063512}.

\bibitem{oyaizu2008nonlinear}
H.~Oyaizu, M.~Lima and W.~Hu, \emph{Nonlinear evolution of f (r) cosmologies.
  ii. power spectrum}, {\emph{Physical Review D} {\bf 78} (2008) 123524}.

\bibitem{Capozziello:2011gm}
S.~Capozziello, M.~De~Laurentis, I.~De~Martino, M.~Formisano and S.~D.
  Odintsov, \emph{{Jeans analysis of self-gravitating systems in
  f(R)-gravity}},
  \href{http://dx.doi.org/10.1103/PhysRevD.85.044022}{\emph{Phys. Rev.} {\bf
  D85} (2012) 044022}, [\href{http://arxiv.org/abs/1112.0761}{{\tt
  1112.0761}}].

\bibitem{PhysRevD.92.044009}
M.~Cataneo, D.~Rapetti, F.~Schmidt, A.~B. Mantz, S.~W. Allen, D.~E. Applegate
  et~al., \emph{New constraints on $f(r)$ gravity from clusters of galaxies},
  \href{http://dx.doi.org/10.1103/PhysRevD.92.044009}{\emph{Phys. Rev. D} {\bf
  92} (Aug, 2015) 044009}.

\bibitem{PhysRevD.93.084016}
A.~de~la Cruz-Dombriz, P.~K.~S. Dunsby, S.~Kandhai and D.~S\'aez-G\'omez,
  \emph{Theoretical and observational constraints of viable $f(r)$ theories of
  gravity}, \href{http://dx.doi.org/10.1103/PhysRevD.93.084016}{\emph{Phys.
  Rev. D} {\bf 93} (Apr, 2016) 084016}.

\bibitem{Pavlidou:2014aia}
V.~Pavlidou, N.~Tetradis and T.~N. Tomaras, \emph{{Constraining Dark Energy
  through the Stability of Cosmic Structures}},
  \href{http://dx.doi.org/10.1088/1475-7516/2014/05/017}{\emph{JCAP} {\bf 1405}
  (2014) 017}, [\href{http://arxiv.org/abs/1401.3742}{{\tt 1401.3742}}].

\bibitem{Pavlidou:2013zha}
V.~Pavlidou and T.~N. Tomaras, \emph{{Where the world stands still: turnaround
  as a strong test of $\Lambda$ CDM cosmology}},
  \href{http://dx.doi.org/10.1088/1475-7516/2014/09/020}{\emph{JCAP} {\bf 1409}
  (2014) 020}, [\href{http://arxiv.org/abs/1310.1920}{{\tt 1310.1920}}].

\bibitem{Tanoglidis:2014lea}
D.~Tanoglidis, V.~Pavlidou and T.~Tomaras, \emph{{Testing $\Lambda$CDM
  cosmology at turnaround: where to look for violations of the bound?}},
  \href{http://dx.doi.org/10.1088/1475-7516/2015/12/060}{\emph{JCAP} {\bf 1512}
  (2015) 060}, [\href{http://arxiv.org/abs/1412.6671}{{\tt 1412.6671}}].

\bibitem{Tanoglidis:2016lrj}
D.~Tanoglidis, V.~Pavlidou and T.~Tomaras, \emph{{Turnaround overdensity as a
  cosmological observable: the case for a local measurement of $\Lambda$}},
  {\emph{arXiv} (2016) }, [\href{http://arxiv.org/abs/1601.03740}{{\tt
  1601.03740}}].

\bibitem{nojiri2018effects}
S.~Nojiri, S.~D. Odintsov and V.~Faraoni, \emph{Effects of modified gravity on
  the turnaround radius in cosmology}, {\emph{arXiv preprint arXiv:1806.01966}
  (2018) }.

\bibitem{khoury2004chameleon}
J.~Khoury and A.~Weltman, \emph{Chameleon cosmology}, {\emph{Physical Review D}
  {\bf 69} (2004) 044026}.

\bibitem{lee2015bound}
J.~Lee, S.~Kim and S.-C. Rey, \emph{{A Bound Violation on the Galaxy Group
  Scale: the Turn-Around Radius of NGC 5353/4}}, {\emph{The Astrophysical
  Journal} {\bf 815} (2015) 43}.

\bibitem{2017arXiv170906903L}
J.~{Lee}, \emph{{Estimating the Turn-Around Radii of Six Isolated Galaxy Groups
  in the Local Universe}}, {\emph{ArXiv e-prints} (Sept., 2017) },
  [\href{http://arxiv.org/abs/1709.06903}{{\tt 1709.06903}}].

\bibitem{Bhattacharya2017cosmic}
S.~Bhattacharya and T.~N. Tomaras, \emph{Cosmic structure sizes in generic dark
  energy models}, {\emph{The European Physical Journal C} {\bf 77} (2017) 526}.

\bibitem{lee2017effect}
J.~Lee and B.~Li, \emph{The effect of modified gravity on the odds of the bound
  violations of the turn-around radii}, {\emph{The Astrophysical Journal} {\bf
  842} (2017) 2}.

\bibitem{bhattacharya2017maximum}
S.~Bhattacharya, K.~F. Dialektopoulos, A.~E. Romano, C.~Skordis and T.~N.
  Tomaras, \emph{The maximum sizes of large scale structures in alternative
  theories of gravity}, {\emph{Journal of Cosmology and Astroparticle Physics}
  {\bf 2017} (2017) 018}.

\bibitem{lopes2018turnaround}
R.~C. Lopes, R.~Voivodic, L.~R. Abramo and L.~Sodr{\'e}~Jr, \emph{Turnaround
  radius in $ f (r) $ model}, {\emph{arXiv preprint arXiv:1805.09918} (2018) }.

\bibitem{tanoglidis2015testing}
D.~Tanoglidis, V.~Pavlidou and T.~Tomaras, \emph{Testing $\lambda$cdm cosmology
  at turnaround: where to look for violations of the bound?}, {\emph{Journal of
  Cosmology and Astroparticle Physics} {\bf 2015} (2015) 060}.

\bibitem{sotiriou2010f}
T.~P. Sotiriou and V.~Faraoni, \emph{f (r) theories of gravity}, {\emph{Reviews
  of Modern Physics} {\bf 82} (2010) 451}.

\bibitem{beraldo2013testing}
L.~J. Beraldo~e Silva, M.~Lima and L.~Sodr{\'e}, \emph{Testing phenomenological
  and theoretical models of dark matter density profiles with galaxy clusters},
  {\emph{Monthly Notices of the Royal Astronomical Society} {\bf 436} (2013)
  2616--2624}.

\bibitem{Navarro:1995iw}
J.~F. Navarro, C.~S. Frenk and S.~D.~M. White, \emph{{The Structure of cold
  dark matter halos}}, \href{http://dx.doi.org/10.1086/177173}{\emph{Astrophys.
  J.} {\bf 462} (1996) 563--575},
  [\href{http://arxiv.org/abs/astro-ph/9508025}{{\tt astro-ph/9508025}}].

\bibitem{navarro1997universal}
J.~F. Navarro, C.~S. Frenk and S.~D. White, \emph{A universal density profile
  from hierarchical clustering}, {\emph{The Astrophysical Journal} {\bf 490}
  (1997) 493}.

\bibitem{bullock2001profiles}
J.~S. Bullock, T.~S. Kolatt, Y.~Sigad, R.~S. Somerville, A.~V. Kravtsov, A.~A.
  Klypin et~al., \emph{Profiles of dark haloes: evolution, scatter and
  environment}, {\emph{Monthly Notices of the Royal Astronomical Society} {\bf
  321} (2001) 559--575}.

\bibitem{cooray2002halo}
A.~Cooray and R.~Sheth, \emph{Halo models of large scale structure},
  {\emph{Physics Reports} {\bf 372} (2002) 1--129}.

\bibitem{hayashi2008understanding}
E.~Hayashi and S.~D. White, \emph{Understanding the halo-mass and galaxy-mass
  cross-correlation functions}, {\emph{Monthly Notices of the Royal
  Astronomical Society} {\bf 388} (2008) 2--14}.

\bibitem{schmidt2009nonlinear}
F.~Schmidt, M.~Lima, H.~Oyaizu and W.~Hu, \emph{Nonlinear evolution of f (r)
  cosmologies. iii. halo statistics}, {\emph{Physical Review D} {\bf 79} (2009)
  083518}.

\bibitem{tinker2010large}
J.~L. Tinker, B.~E. Robertson, A.~V. Kravtsov, A.~Klypin, M.~S. Warren,
  G.~Yepes et~al., \emph{The large-scale bias of dark matter halos: numerical
  calibration and model tests}, {\emph{The Astrophysical Journal} {\bf 724}
  (2010) 878}.

\bibitem{hojjati2011testing}
A.~Hojjati, L.~Pogosian and G.-B. Zhao, \emph{Testing gravity with camb and
  cosmomc}, {\emph{Journal of Cosmology and Astroparticle Physics} {\bf 2011}
  (2011) 005}.

\bibitem{mo2010galaxy}
H.~Mo, F.~Van~den Bosch and S.~White, \emph{Galaxy formation and evolution}.
\newblock Cambridge University Press, 2010.

\bibitem{Faraoni1}
V.~{Faraoni} and S.~{Capozziello}, \emph{{Beyond Einstein Gravity: A Survey of
  Gravitational Theories for Cosmology and Astrophysics}}.
\newblock 2011.

\bibitem{martino2009spherical}
M.~C. Martino, H.~F. Stabenau and R.~K. Sheth, \emph{Spherical collapse and
  cluster counts in modified gravity models}, {\emph{Physical Review D} {\bf
  79} (2009) 084013}.

\bibitem{cupani2008mass}
G.~Cupani, M.~Mezzetti and F.~Mardirossian, \emph{Mass estimation in the outer
  non-equilibrium region of galaxy clusters}, {\emph{Monthly Notices of the
  Royal Astronomical Society} {\bf 390} (2008) 645--654}.

\bibitem{eke1996cluster}
V.~R. Eke, S.~Cole and C.~S. Frenk, \emph{Cluster evolution as a diagnostic for
  $\omega$}, {\emph{Monthly Notices of the Royal Astronomical Society} {\bf
  282} (1996) 263--280}.

\bibitem{kashibadze2018cosmic}
O.~G. Kashibadze and I.~D. Karachentsev, \emph{Cosmic flow around local massive
  galaxies}, {\emph{Astronomy \& Astrophysics} {\bf 609} (2018) A11}.

\bibitem{burrage2018tests}
C.~Burrage and J.~Sakstein, \emph{Tests of chameleon gravity}, {\emph{Living
  reviews in relativity} {\bf 21} (2018) 1}.

\bibitem{RinesDiaferio}
K.~{Rines} and A.~{Diaferio}, \emph{Cirs: Cluster infall regions in the sloan
  digital sky survey. i. infall patterns and mass profiles},
  \href{http://dx.doi.org/10.1086/506017}{\emph{The Astronomical Journal} {\bf
  132} (Sept., 2006) 1275--1297},
  [\href{http://arxiv.org/abs/astro-ph/0602032}{{\tt astro-ph/0602032}}].

\bibitem{Snaith}
O.~N. {Snaith}, J.~{Bailin}, A.~{Knebe}, G.~{Stinson}, J.~{Wadsley} and
  H.~{Couchman}, \emph{{Haloes at the ragged edge: the importance of the
  splashback radius}},
  \href{http://dx.doi.org/10.1093/mnras/stx2138}{\emph{Monthly Notices of the
  Royal Astronomical Society} {\bf 472} (Dec., 2017) 2694--2712},
  [\href{http://arxiv.org/abs/1708.06181}{{\tt 1708.06181}}].

\bibitem{Chang}
C.~{Chang}, E.~{Baxter}, B.~{Jain}, C.~{S{\'a}nchez}, S.~{Adhikari}, T.~N.
  {Varga} et~al., \emph{{The Splashback Feature around DES Galaxy Clusters:
  Galaxy Density and Weak Lensing Profiles}},
  \href{http://dx.doi.org/10.3847/1538-4357/aad5e7}{\emph{The Astrophysical
  Journal} {\bf 864} (Sept., 2018) 83},
  [\href{http://arxiv.org/abs/1710.06808}{{\tt 1710.06808}}].

\bibitem{More}
S.~{More}, B.~{Diemer} and A.~V. {Kravtsov}, \emph{{The Splashback Radius as a
  Physical Halo Boundary and the Growth of Halo Mass}},
  \href{http://dx.doi.org/10.1088/0004-637X/810/1/36}{\emph{The Astrophysical
  Journal} {\bf 810} (Sept., 2015) 36},
  [\href{http://arxiv.org/abs/1504.05591}{{\tt 1504.05591}}].

\bibitem{Okumura}
T.~{Okumura}, T.~{Nishimichi}, K.~{Umetsu} and K.~{Osato}, \emph{{Splashback
  radius of nonspherical dark matter halos from cosmic density and velocity
  fields}}, \href{http://dx.doi.org/10.1103/PhysRevD.98.023523}{\emph{Physical
  Review D} {\bf 98} (July, 2018) 023523},
  [\href{http://arxiv.org/abs/1807.02669}{{\tt 1807.02669}}].

\bibitem{Adhikari}
S.~{Adhikari}, J.~{Sakstein}, B.~{Jain}, N.~{Dalal} and B.~{Li},
  \emph{{Splashback in galaxy clusters as a probe of cosmic expansion and
  gravity}}, {\emph{ArXiv e-prints} (June, 2018) },
  [\href{http://arxiv.org/abs/1806.04302}{{\tt 1806.04302}}].

\bibitem{DESI}
{DESI Collaboration}, A.~{Aghamousa}, J.~{Aguilar}, S.~{Ahlen}, S.~{Alam},
  L.~E. {Allen} et~al., \emph{{The DESI Experiment Part I: Science,Targeting,
  and Survey Design}}, {\emph{ArXiv e-prints} (Oct., 2016) },
  [\href{http://arxiv.org/abs/1611.00036}{{\tt 1611.00036}}].

\bibitem{JPAS}
N.~{Benitez}, R.~{Dupke}, M.~{Moles}, L.~{Sodre}, J.~{Cenarro},
  A.~{Marin-Franch} et~al., \emph{{J-PAS: The Javalambre-Physics of the
  Accelerated Universe Astrophysical Survey}}, {\emph{ArXiv e-prints} (Mar.,
  2014) }, [\href{http://arxiv.org/abs/1403.5237}{{\tt 1403.5237}}].

\end{thebibliography}\endgroup
\end{document}